\documentclass[a4paper,12pt]{scrartcl}

 \usepackage[]{graphicx}\usepackage[]{color}
\usepackage{comment}

\usepackage{alltt}
\usepackage[left= 2.5cm,right = 2.5cm, bottom = 3cm,top=2cm]{geometry}
\usepackage[utf8]{inputenc} 
\usepackage{setspace}
\usepackage{textcomp}
\usepackage[flushleft]{threeparttable}
\usepackage{nth}

\usepackage{amsmath}
\usepackage{amsfonts}
\usepackage{amssymb}
\usepackage{graphics}
\usepackage{graphicx}
\usepackage{epstopdf}
\usepackage[nottoc]{tocbibind}
\usepackage[table]{xcolor} 
\usepackage{setspace}
\usepackage{url}
\usepackage{multirow}
\usepackage[authoryear]{natbib}
\usepackage{hyperref}
\usepackage{cleveref} 
\newcommand{\GG}[1]{}

\newlength{\minuslength}
\usepackage{subfig}

\usepackage{bm}

\usepackage{booktabs}

\usepackage{acronym} 
\usepackage{here} 

\usepackage{titling}

\title{Sharpening Identification in Large Structural VARs Using Narrative Restrictions\thanks{We thank Boris Blagov, Max Breitenlechner, Robert Czudaj and Martin Geiger for their valuable feedback and the participants at the workshop Macroeconometrics in Salerno at the University of Salerno, Italy and at the workshop on Global Challenges for Emerging
Markets in Times of Disruptions and Increasing Uncertainty of the Universidad de los Andes in Bogot\'a, Colombia. Jan Pr\"user gratefully
acknowledges the support of the German Research Foundation (DFG, 561170677).}} 

\author{Lukas Berend$^{a}$ and Jan Pr\"user$^{b}$
\\[0.2cm]
$^{a}${\small FernUniversit\"at in Hagen\thanks{Corresponding Author: Fakult\"at f\"ur Wirtschaftswissenschaft, 58097 Hagen, Germany, e-mail: \texttt{lukas.berend@fernuni-hagen.de} }}
$^{b}${\small TU Dortmund\thanks{ Fakult\"at Statistik, 44221 Dortmund, Germany, e-mail: \texttt{prueser@statistik.tu-dortmund.de\,}} \hspace{0.1cm} 
} }

\begin{document}

\date{\today}
\maketitle
\begin{abstract}
	\begin{singlespace}
	\noindent
We propose a high-dimensional structural vector autoregression framework with a factor structure in the error terms that accommodates a large number of linear inequality restrictions on both impact impulse responses and structural shocks. Our framework extends recent advances in large sign-restricted VARs by allowing narrative restrictions to be imposed directly through constraints on structural shocks via prior distributions, thereby sharpening identification and enhancing the economic interpretability of the structural shocks. To estimate the model, we develop a computationally efficient sampling algorithm that scales well with both model dimension and the number of imposed restrictions, while avoiding the low acceptance-rate problems associated with existing rejection-based approaches. We apply our methodology to a large-scale structural VAR model of the U.S. economy, identifying ten structural shocks and tracing their dynamic effects across thirty-nine macroeconomic and financial variables. The empirical application demonstrates that the incorporation of narrative restrictions improves structural identification in high-dimensional settings by reducing the uncertainty surrounding impulse response functions and facilitating a clearer economic interpretation of the identified structural shocks.

	\end{singlespace}
\end{abstract}
\bigskip
 
 
	 \bigskip

\thispagestyle{empty} 
\newpage

\pagenumbering{arabic}
\section{Introduction}

\cite{uhlig2005effects} popularized the use of sign restrictions on impulse response functions as a flexible and theory-consistent approach to identifying structural shocks in Vector Autoregression (VAR) models. Rather than relying on often implausible exact zero restrictions, sign restrictions impose economically motivated inequality constraints on the responses of variables to structural disturbances. More recently, sign and ranking restrictions have become increasingly popular in large-dimensional VARs, as they provide a tractable and economically interpretable framework for identifying structural shocks in information-rich environments; see \cite{Korobilis2022}, \cite{arias2025large}, and \cite{chan2025rankingrestrictions}. 

Building on the sign restriction framework of \cite{uhlig2005effects}, \cite{antolin2018narrative} introduce narrative restrictions, which incorporate prior information about specific historical episodes---such as well-documented policy interventions, financial crises, or geopolitical events---to sharpen set identification.\footnote{The sign restrictions on shocks over distinct time periods may be interpreted as a more agnostic framework than impact sign restrictions, as noted in \cite{antolin2018narrative}. Nonetheless, such restrictions must remain grounded in a plausible historical narrative.} By conditioning structural shocks on specific periods, narrative restrictions allow researchers to embed historically grounded information directly into the identification scheme, thereby enhancing both interpretability and credibility. As a result, narrative restrictions have become an increasingly important tool in empirical macroeconomics.\footnote{Important examples include \cite{zeev2018can}, \cite{altavilla2019loan}, \cite{furlanetto2019immigration}, \cite{cheng2020revisiting}, \cite{kilian2020does}, \cite{kilian2022oil}, \cite{laumer2020government}, \cite{redl2020uncertainty}, \cite{zhou2020refining}, \cite{antolin2021structural}, \cite{caggiano2021financial}, \cite{larsen2021components}, \cite{ludvigson2021uncertainty}, \cite{maffei2021does}, \cite{berger2022unified}, \cite{fanelli2022sovereign}, \cite{inoue2022joint}, \cite{badinger2023measuring}, \cite{berthold2023macroeconomic}, \cite{caggiano2023global}, \cite{conti2023bank}, \cite{harrison2023structural}, \cite{herwartz2023point}, \cite{neri2023long}, \cite{reichlin2023monetary}, \cite{ascari2023endogenous}, \cite{boer2024energy}, and \cite{ruth2023monetary}.} 

Despite their growing empirical relevance, the application of narrative restrictions has so far been confined to relatively small-dimensional VARs. The reason is fundamentally computational. Existing implementations typically rely on pseudo-likelihood-based accept-and-reject algorithms; see, among others, \cite{antolin2018narrative} and \cite{giacomini2021identification}. While effective in low-dimensional settings, these procedures scale poorly as model dimension and the number of restrictions increase, since posterior simulation depends on increasingly unlikely acceptance events. In practice, this acceptance-rate collapse renders narrative restrictions computationally infeasible in information-rich environments and severely limits their use in large structural VARs. 

This limitation is particularly important given the growing prominence of large Bayesian VARs in macroeconomics following the seminal contribution of \cite{banbura2010large}. By incorporating rich information sets, large VARs mitigate omitted variable bias and improve structural identification. This is especially relevant because insufficient information sets may render VARs non-fundamental and thereby prevent the correct recovery of structural shocks, as emphasized by \cite{hansen2019two} and \cite{lippi1993dynamic,lippi1994var}. Building on this insight, a rapidly expanding literature has developed methods for structural inference in high-dimensional systems; see, among others, \cite{carriero2009forecasting}, \cite{giannone2015prior}, \cite{jarocinski2017granger}, \cite{huber2019adaptive}, and \cite{chan2024large}. Yet despite these advances, narrative restrictions have remained largely absent from this literature due to the computational bottlenecks of existing approaches. 

This paper resolves this limitation by developing a high-dimensional structural VAR framework that accommodates large sets of linear inequality restrictions on both impact responses and structural shocks. Our central contribution is to make narrative restrictions computationally feasible in high-dimensional settings.

Our framework builds on recent large VAR models that assume a factor structure in the reduced-form errors, with latent factors interpreted as structural shocks; see, among others, \cite{Korobilis2022}, \cite{korobilis2025exploring}, \cite{chan2022large}, \cite{hauzenberger2022bayesian}, \cite{banbura2023drives}, \cite{gambetti2023agreed}, \cite{pfarrhofer2024high}, \cite{pruser2024large}, \cite{chan2025large}, \cite{bobeica2025beware}, and \cite{clark2025nonparametric}. This factor structure offers several important advantages. First, it enables equation-by-equation estimation, making posterior inference computationally feasible even in large systems. Second, it naturally accommodates settings in which the number of economically relevant structural shocks is substantially smaller than the number of observed variables. Third, and most importantly for our purposes, the latent factors admit a direct interpretation as structural shocks, thereby providing a natural framework for imposing restrictions directly on these shocks. Unlike existing approaches such as \cite{antolin2018narrative} and \cite{giacomini2021identification}, which impose narrative restrictions through pseudo-likelihood-based accept-and-reject procedures, we incorporate these restrictions directly through truncated prior distributions. This eliminates the acceptance-rate bottleneck and yields posterior simulation with acceptance probability equal to one. The resulting computational gains are substantial and allow the approach to scale efficiently with both model dimension and the number of restrictions. Consequently, our framework makes it feasible to jointly impose large sets of impact, ranking, and narrative shock restrictions in settings where existing methods become prohibitively slow or computationally infeasible.

To our knowledge, this paper provides the first application of narrative restrictions in a genuinely data-rich environment. We demonstrate the practical relevance of the proposed framework using a 39-variable structural VAR for the U.S. economy. Building on \cite{arias2025large} and \cite{chan2025rankingrestrictions}, we identify ten structural shocks using a combination of impact sign restrictions, ranking restrictions, and narrative shock restrictions. Specifically, our framework enables us to identify financial, demand, investment, monetary policy, fiscal, technology, labor supply, wage bargaining, oil supply, and consumer sentiment shocks. We show that narrative restrictions materially sharpen identification and improve the interpretability of structurally distinct disturbances. More broadly, our results demonstrate that large-scale structural VARs with tractable narrative restrictions provide a practical and empirically powerful framework for macroeconomic analysis.

The remainder of the paper is organized as follows. Section \ref{emp_framework} presents the econometric framework. Section \ref{section_emp_application} contains the empirical application and discusses the identification of the ten structural shocks. Section 4 concludes.

\section{A Large Structural VAR with Sign and Narrative Restrictions} \label{emp_framework}
First, we lay out the model framework and then discuss how we implement sign and narrative restrictions using truncated prior distributions.
\subsection{ A large structural VAR with factor structure}
Let $\boldsymbol{y}_t = (y_{1,t}, \dots, y_{n,t})'$ represent an $n \times 1$ vector of endogenous variables at time $t$. The model can be expressed as:
\begin{equation}
\boldsymbol{y}_t = \boldsymbol{b}_0 + \boldsymbol{B}_1 \boldsymbol{y}_{t-1} + \cdots + \boldsymbol{B}_p \boldsymbol{y}_{t-p} + \boldsymbol{u}_t,
\end{equation}
where the error term $\boldsymbol{u}_t$ is decomposed as 
\begin{equation}
\boldsymbol{u}_t = \boldsymbol{L} \boldsymbol{f}_t + \boldsymbol{v}_t,
\end{equation}
with $\boldsymbol{v}_t \sim N(\boldsymbol{0}, \boldsymbol{\Sigma})$, where $\boldsymbol{\Sigma} = \text{diag}(\sigma_1^2, \dots, \sigma_n^2)$, and $\boldsymbol{f}_t$ is an $r \times 1$ vector of factors such that $\boldsymbol{f}_t \sim N(\boldsymbol{0}, \boldsymbol{I})$. Concisely, the system can be reformulated as:
\begin{equation}
\boldsymbol{y}_t = (\boldsymbol{I}_n \otimes \boldsymbol{x}'_t) \boldsymbol{\beta} + \boldsymbol{L} \boldsymbol{f}_t + \boldsymbol{v}_t,
\end{equation}
where $\boldsymbol{I}_n$ is the identity matrix of dimension $n$, $\otimes$ represents the Kronecker product, and $\boldsymbol{\beta} = \text{vec}([\boldsymbol{b}_0, \boldsymbol{B}_1, \dots, \boldsymbol{B}_p]')$. The vector $\boldsymbol{x}_t = (1, \boldsymbol{y}'_{t-1}, \dots, \boldsymbol{y}'_{t-p})'$ of dimension $k = 1 + np$ contains an intercept and lagged values. The idiosyncratic component $\boldsymbol{v}_t$ accounts for measurement error or other idiosyncratic noise. In contrast, the $r$ latent factors, $\boldsymbol{f}_t$ impact multiple variables in the system and hence have the interpretation as structural shocks.
More formally we multiply the model with the generalized inverse of $\boldsymbol{L}$ to obtain:
\begin{equation}
\boldsymbol{A} \boldsymbol{y}_t = \boldsymbol{B} \boldsymbol{x}_t + \boldsymbol{f}_t + \boldsymbol{A} \boldsymbol{v}_t,
\end{equation}
where $\boldsymbol{A} = (\boldsymbol{L}' \boldsymbol{L})^{-1} \boldsymbol{L}'$ and $\boldsymbol{B} = (\boldsymbol{A} \boldsymbol{b}_0, \boldsymbol{A} \boldsymbol{B}_1, \dots, \boldsymbol{A} \boldsymbol{B}_p)$. Given that $\boldsymbol{v}_t$ is uncorrelated noise, the Central Limit Theorem (see \cite{bai2003}) implies that $\boldsymbol{A} \boldsymbol{v}_t \to 0$ as $n \to \infty$. Hence, we can write
\begin{equation}
\boldsymbol{f}_t \approx \boldsymbol{A} \boldsymbol{y}_t - \boldsymbol{B} \boldsymbol{x}_t.
\end{equation}
Thus, $\boldsymbol{v}_t$ is treated as noise shocks without structural meaning, while $\boldsymbol{f}_t$ provides a projection of structural shocks in $\mathbb{R}^r$. Consequently, this formulation supports structural analysis using standard methods, including impulse response functions (e.g., \cite{Froni2019}, \cite{Korobilis2022}, \cite{chan2022large}).
Under the assumption of uncorrelated $\boldsymbol{f}_t$ and $\boldsymbol{v}_t$, the conditional variance of $\boldsymbol{u}_t$ is given by:
\begin{equation}
\text{Var}(\boldsymbol{u}_t | \boldsymbol{\Sigma}, \boldsymbol{L}) = \boldsymbol{L} \boldsymbol{L}' + \boldsymbol{\Sigma}.
\end{equation}
To separately identify the common and idiosyncratic components, we adopt the condition $r \leq (n-1)/2$ from \cite{anderson1956statistical}. Under this restriction, for any observationally equivalent model $(\boldsymbol{L},\boldsymbol{\Sigma})$, it holds that $\boldsymbol{L}  \boldsymbol{L}' = \boldsymbol{L}^*  \boldsymbol{L}^{*'}$ and $\boldsymbol{\Sigma} = \boldsymbol{\Sigma}^*$.
However, without additional restrictions, $\boldsymbol{L}$ is not identified. Any orthogonal matrix $\boldsymbol{Q} \in \mathcal{O}(r)$, where $\mathcal{O}(r) = \{ \boldsymbol{Q} \in \mathbb{R}^{r \times r} : \boldsymbol{Q} \boldsymbol{Q}' = \boldsymbol{I}_m\}$, generates an equivalent model via the transformation $\tilde{\boldsymbol{L}} \tilde{\boldsymbol{f}}_t = \boldsymbol{L} \boldsymbol{Q} \boldsymbol{Q}' \boldsymbol{f}_t$. In this paper, we achieve set identification by placing inequality restrictions on $\boldsymbol{L}$ and $\boldsymbol{f}_{t}$.

\subsection{Linear inequality restrictions via prior distributions}

We consider inequality restrictions on $\boldsymbol{L}$ and $\boldsymbol{f}_{t}$.\footnote{In many cases we have a strong consensus in economic theory about the
signs of impulse responses at impact but not at longer horizons (see, e.g. \cite{canova2011business}).} We can express these inequality restrictions as follows:
\begin{align}
\ell_i^{L} &< \boldsymbol{R}_{i}^{L} \boldsymbol{l}_{i'}< \upsilon_i^{L},\\
\ell_{\tilde{t}}^{f} &< \boldsymbol{R}_{\tilde{t}}^{f} \boldsymbol{f}_{\tilde{t}}< \upsilon_{\tilde{t}}^{f},
\end{align}
where $\boldsymbol{l}_i$ denote the elements of $\boldsymbol{L}$ in the $i-$th equation and $\boldsymbol{R}_{i}^{L}$ is $r_i^L\times r$, 
$\boldsymbol{R}_{\tilde{t}}^{f}$ is $r_{\tilde{t}}^f\times r$ and $\tilde{t}\in t=1,\dots,T$. We allow elements in the lower and upper bound to be $-\infty$ or $\infty$ to indicate that some inequity restrictions are single-sided. In many applications $\boldsymbol{R}_i^{L}$ and $\boldsymbol{R}_{\tilde{t}}^{f}$ are equal to $\boldsymbol{I}_r$. The first set of inequality restrictions can be used to implement sign restrictions on $\boldsymbol{L}$ as in \cite{Korobilis2022}. The second set of inequality restriction can be used to implement sign restrictions on the structural shocks $\boldsymbol{f}_{t}$ at specific time periods as in \cite{antolin2018narrative}.\footnote{Also, combining both restrictions can be used to implement magnitude restrictions on the product of both factors loadings and structural shocks similar to \cite{Badinger2023}. In particular, \cite{Badinger2023} impose that one particular shock explains more then half of the unexplained movement of a certain variable at a specific period.} Following the discussion in \cite{giacomini2022narrative, giacomini2022narrativerejoinder} and \cite{plagborg2022discussion}, we note that imposing shock sign restrictions for specific time periods displays a more agnostic approach than magnitude restrictions.


We implement our inequality restrictions using  truncated Gaussian prior distributions. In particular we assume
\begin{align}
\boldsymbol{l}_i &\sim N(\boldsymbol{l}_{0,i},\boldsymbol{V}_{\boldsymbol{l}_{i}})\boldsymbol{1} (\ell_i^{L} < \boldsymbol{R}_i^{L} \boldsymbol{l}_i'< \upsilon_i^{L})\\
\boldsymbol{f}_t &\sim N(\boldsymbol{0}, \boldsymbol{I})\boldsymbol{1} (\ell_{t}^{f} < \boldsymbol{R}_{t}^{f} \boldsymbol{f}_{t}< \upsilon_{t}^{f}),
\end{align}
where $\tilde{T}$ denotes the set of periods in which a restrictions is specified. In the Online Appendix, we complete our model specification by describing the remaining prior distributions and outline the Gibbs sampler to estimate the model.\footnote{To address over parameterization concerns we use the horseshoe prior, following \cite{carvalho2010horseshoe} and \cite{Korobilis2022}.}

\section{Empirical Application: Structural Shocks in the U.S. Economy}\label{section_emp_application}

In this section, we estimate a large-scale SVAR comprising 39 macroeconomic and financial variables and identify ten structural shocks for the U.S. economy. The empirical application serves to illustrate the effectiveness of our proposed algorithm for jointly imposing impact, ranking, and shock sign restrictions in a high-dimensional setting. In particular, we demonstrate that augmenting standard identification schemes with narrative sign restrictions sharpens the identification of structural shocks by and tightening inference on impulse responses.

Our specification builds on \cite{arias2025large} and \cite{chan2025rankingrestrictions}, who combine sign and ranking restrictions to identify ten structural shocks.\footnote{\cite{arias2025large} extend the framework of \cite{chan2025rankingrestrictions} by adding an oil price shock and a consumer confidence shock.} Relative to this literature, we show that incorporating narrative sign restrictions further sharpens identification and leads to more precise and interpretable estimates of the dynamic effects.

The model includes the same core variables as in \cite{arias2025large}, augmented by corporate real liquid assets, corporate credit spreads, the excess bond premium of \cite{gilchrist2012credit}, and the economic policy uncertainty (EPU) index of \cite{baker2016measuring}.\footnote{Appendix \ref{data_set} provides a detailed description of all 39 variables, including data sources and transformations.} Owing to data availability and distortions related to the COVID-19 period, the sample is restricted to 1977Q1--2019Q4. The VAR is estimated with a lag length of $p = 5$ and includes a constant term.\footnote{In the Gibbs sampler, we discard the first 500 draws as burn-in and retain every \nth{10} draw from a total of 5{,}000 draws as posterior draws.}

For identification, we adopt the sign and ranking restrictions of \cite{arias2025large}, with several modifications.\footnote{We assume consumer sentiment to increase following a consumer confidence shock. Moreover, we reinterpret the oil price shock as an oil supply shock, drop the sign restriction on unemployment, and instead impose that industrial production increases following a positive oil supply shock. This adjustment facilitates the inclusion of narrative restrictions based on \cite{antolin2018narrative}.} In addition, we impose that a financial shock tightens credit conditions, reflected in increases in liquidity constraints, the GZ spread, and the excess bond premium. Consistent with \cite{brianti2025financial}, we further assume that liquidity increases alongside the excess bond premium (see Table~\ref{tab:sign-rank-restrictions}). We also impose that the GZ spread declines in response to a positive demand shock \citep{de2026deflationary}.

Regarding ranking restrictions, we follow \cite{arias2025large}. Interpreting financial shocks as disruptions to credit conditions, we require the GZ spread to respond more strongly than the EPU index.\footnote{To distinguish adverse financial shocks from negative demand and consumer confidence shocks, we impose that the GZ spread exhibits the largest absolute response among these disturbances.} In total, as summarized in Table \ref{tab:sign-rank-restrictions}, we impose 139 sign and ranking restrictions to identify demand, investment, financial, monetary, government spending, technology, labor, wage bargaining, oil supply, and consumer sentiment shocks.

\begin{table}[h]
\centering
\scriptsize
\renewcommand{\arraystretch}{0.85} 
\setlength{\tabcolsep}{2.5pt}       
\begin{tabular}{lcccccccccc}
\hline
\textbf{Variable} & \textbf{Dem} & \textbf{Inv} & \textbf{Fin} & \textbf{Mon} & \textbf{Gov} & \textbf{Tec} & \textbf{Lab} & \textbf{Wag} & \textbf{Oil} & \textbf{Con}  \\
\hline
GDP & +1 & +1 & -1 & -1 & +1 & +1 & +1 & +1 & +1 & +1 \\
PCE & 0 & 0 & 0 & 0 & 0 & +1 & 0 & 0 & +1 & +1  \\
Residential investment & 0 & 0 & 0 & 0 & 0 & 0 & 0 & 0 & 0 & +1 \\
Nonresidential investment & 0 & +1 & 0 & 0 & 0 & +1 & 0 & 0 & +1 & +1 \\
Exports & 0 & 0 & 0 & 0 & 0 & 0 & 0 & 0 & 0 & 0 \\
Imports & 0 & 0 & 0 & 0 & 0 & 0 & 0 & 0 & 0 & 0 \\
Government spending & 0 & 0 & 0 & 0 & +1 & 0 & 0 & 0 & 0 & 0 \\
Fed. budget surplus/deficit & 0 & 0 & 0 & 0 & -1 & 0 & 0 & 0 & 0 & 0 \\
Fed. tax receipts & 0 & 0 & 0 & 0 & +1 & 0 & 0 & 0 & 0 & 0 \\
GDP deflator & +1 & +1 & -1 & -1 & +1 & -1 & -1 & -1 & -1 & +1 \\
PCE index & +1 & +1 & -1 & -1 & +1 & -1 & -1 & -1 & -1 & +1 \\
PCE index less F\&E & +1 & +1 & -1 & -1 & +1 & -1 & -1 & -1 & -1 & +1 \\
CPI index & +1 & +1 & -1 & -1 & +1 & -1 & -1 & -1 & -1 & +1 \\
CPI index less F\&E & +1 & +1 & -1 & -1 & +1 & -1 & -1 & -1 & -1 & +1 \\
Hourly wage & 0 & 0 & 0 & 0 & 0 & +1 & -1 & -1 & +1 & 0 \\
Labor productivity & 0 & 0 & 0 & 0 & 0 & +1 & 0 & 0 & +1 & 0 \\
Utilization-adjusted TFP & 0 & 0 & 0 & 0 & 0 & +1 & 0 & 0 & +1 & 0  \\
Employment & 0 & 0 & 0 & -1 & 0 & 0 & -1 & 0 & 0 & 0  \\
Unemployment rate & -1 & -1 & +1 & +1 & -1 & -1 & +1 & -1 & 0 & +1 \\
Industrial production & +1 & +1 & -1 & -1 & 0 & 0 & 0 & 0 & 1 & 0 \\
Capacity utilization & +1 & +1 & -1 & -1 & 0 & 0 & 0 & 0 & 0 & 0 \\
Housing starts & 0 & 0 & 0 & 0 & 0 & 0 & 0 & 0 & 0 & 0 \\
Disposable income & 0 & 0 & 0 & 0 & 0 & 0 & 0 & 0 & 0 & 0 \\
Consumer sentiment & 0 & 0 & 0 & 0 & 0 & 0 & 0 & 0 & 0 & 1 \\
Fed funds rate & +1 & +1 & -1 & +1 & +1 & 0 & 0 & 0 & 0 & 0 \\
3-month T-bill rate & +1 & +1 & -1 & +1 & +1 & 0 & 0 & 0 & 0 & 0 \\
2-year T-note rate & 0 & 0 & 0 & +1 & 0 & 0 & 0 & 0 & 0 & 0 \\
5-year T-note rate & 0 & 0 & 0 & +1 & 0 & 0 & 0 & 0 & 0 & 0 \\
10-year T-note rate & 0 & 0 & 0 & +1 & 0 & 0 & 0 & 0 & 0 & 0\\
Prime rate & +1 & +1 & -1 & +1 & +1 & 0 & 0 & 0 & 0 & 0 \\
Aaa corporate bond yield & 0 & 0 & 0 & +1 & 0 & 0 & 0 & 0 & 0 & 0 \\
Baa corporate bond yield & 0 & 0 & 0 & +1 & 0 & 0 & 0 & 0 & 0 & 0 \\
Trade-weighted US index & 0 & 0 & 0 & 0 & 0 & 0 & 0 & 0 & 0 & 0 \\
S\&P 500 & 0 & -1 & -1 & -1 & 0 & 0 & 0 & 0 & 0 & +1 \\
Spot oil price & 0 & 0 & 0 & 0 & 0 & 0 & 0 & 0 & -1 & 0 \\
Liquidity & 0 & 0 & -1 & 0 & 0 & 0 & 0 & 0 & 0 & 0 \\
GZ spread & -1 & 0 & +1 & +1 & 0 & 0 & 0 & 0 & 0 & 0 \\
EBP & 0 & 0 & +1 & +1 & 0 & 0 & 0 & 0 & 0 & 0  \\
EPU & 0 & 0 & 0 & 0 & 0 & 0 & 0 & 0 & 0 & 0  \\ 
\hline
\textbf{Ranking restrictions} & & & & & & & & &  \\
Nonresidential investment/GDP & -1 & +1 & +1 & 0 & 0 & 0 & 0 & 0 & 0 & 0 \\
Government spending/GDP & -1 & -1 & -1 & 0 & +1 & 0 & 0 & 0 & 0 & 0 \\
EPU/GZ spread & 0 & 0 & -1 & 0 & 0 & 0 & 0 & 0 & 0 & 0 \\
\hline
\textbf{(Cumulative) Number of restrictions} & & & & & & & & &  \\
No of restrictions & 15 & 16 & 19 & 21 & 14 & 12 & 9 & 8 & 13 & 12  \\
Cum. No of restrictions & 15 & 31 & 50 & 71 & 85 & 97 & 106 & 114 & 127 & 139  \\
\hline
\end{tabular}

\vspace{0.5em}
\parbox{\textwidth}{\scriptsize
\textit{Notes:} For the sign restrictions, +1 (-1) indicate a positive (negative) response to the respective shock.
For ranking restrictions, +1 (-1) indicate that the variable in the numerator responds stronger (weaker) than the variable in the denominator.
The abbreviations correspond as follows: Dem: demand; Inv: investment; Fin: financial; Mon: monetary policy; Gov: government spending; Tec: technology; Lab: labor supply; Wag: wage bargaining; Oil: oil supply; Con: consumer sentiment.}

\caption{Sign restrictions, ranking restrictions, and identified shocks in the baseline model.}
\label{tab:sign-rank-restrictions}
\end{table}

In our extended model, we augment the above baseline model with a set of narrative sign restrictions. First, we impose a positive demand shock in 2006Q1, reflecting that GDP growth exceeded FOMC expectations and was primarily driven by consumption \citep{de2026deflationary}.\footnote{While \cite{de2026deflationary} attribute the forecast error in GDP predominantly to a demand shock, we restrict only the sign of the demand shock itself.} For financial shocks, we impose positive realizations in 2007Q3, 2008Q3, and 2008Q4. The 2007Q3 restriction captures the tightening in credit conditions following the downgrade of subprime-related assets, while the 2008Q3 and 2008Q4 restrictions reflect the financial turmoil associated with the collapse of Lehman Brothers and the corresponding surge in credit spreads. For oil supply shocks, we impose negative realizations in 1990Q3 (Gulf War), 2002Q4 (Venezuelan unrest), 2003Q1 (Iraq War), and 2011Q1 (Libyan civil war), following \cite{antolin2018narrative}. For government spending shocks, we impose positive realizations in 1980Q1, associated with anticipated increases in U.S. defense spending following the Soviet invasion of Afghanistan, and in 2001Q3, reflecting the fiscal response to the September 11 attacks, in line with \cite{laumer2020government}.

In total, the extended model incorporates 10 narrative restrictions in addition to the 139 sign and ranking restrictions.

\begin{table}[H]
\centering
\renewcommand{\arraystretch}{0.9} 
\begin{tabular}{lcccccccccc}
\hline
\textbf{Quarter} & \textbf{Dem} & \textbf{Inv} & \textbf{Fin} & \textbf{Mon} & \textbf{Gov} & \textbf{Tec} & \textbf{Lab} & \textbf{Wag} & \textbf{Oil} & \textbf{Con} \\
\hline
1980Q1 &  &  &  &  & $>0$ &  &  &  &  &   \\
1990Q3 &  &  &  &  &  &  &  &  & $<0$ &    \\
2001Q3 &  &  &  &  & $>0$ &  &  &  &  &   \\
2002Q4 &  &  &  &  &  &  &  &  & $<0$ &    \\
2003Q1 &  &  &  &  &  &  &  &  & $<0$ &    \\
2006Q1 & $>0$ &  &  &  & &  &  &  &  &    \\
2007Q3 &  &  & $>0$ &  &  &  &  &  &  &  \\
2008Q3 &  &  & $>0$ &  &  &  &  &  &  &   \\
2008Q4 &  &  & $>0$ &  &  &  &  &  &  &   \\
2011Q1 &  &  &  &  &  &  &  &  & $<0$ &    \\
\hline
\end{tabular}

\vspace{0.5em}
\parbox{\textwidth}{\scriptsize
\textit{Notes:} The table reports shock sign restrictions imposed in the extended model.
Entries $>0$ ($<0$) indicate that the corresponding shock is assumed to positive (negative)
in the respective quarter. The abbreviations correspond as follows: Dem: demand; Inv: investment; Fin: financial; Mon: monetary policy; Gov: government spending;
Tec: technology; Lab: labor supply; Wag: wage bargaining; Oil: oil supply; Con: consumer sentiment.}

\caption{Narrative shock sign restrictions in the extended model.}
\label{tab:shock-quarter-restrictions}
\end{table}

\subsection{Empirical Results}

This section presents selected empirical results illustrating how the inclusion of narrative restrictions on structural shocks sharpens identification in the VAR model. Specifically, augmenting standard sign and ranking restrictions with narrative information reduces the size of the admissible set, which is reflected in narrower credible bands of the impulse response functions. To highlight this effect, we compare IRFs obtained from a baseline specification without narrative restrictions to those from a model that incorporates sign, ranking, and narrative restrictions.

\begin{figure}[H]
\centering
\setlength{\tabcolsep}{2pt}
\renewcommand{\arraystretch}{1}

\begin{tabular}{ccc}
GDP & Investment & Imports \\

\includegraphics[
    width=0.32\textwidth,
    trim=1mm 1mm 1mm 1mm,
    clip
]{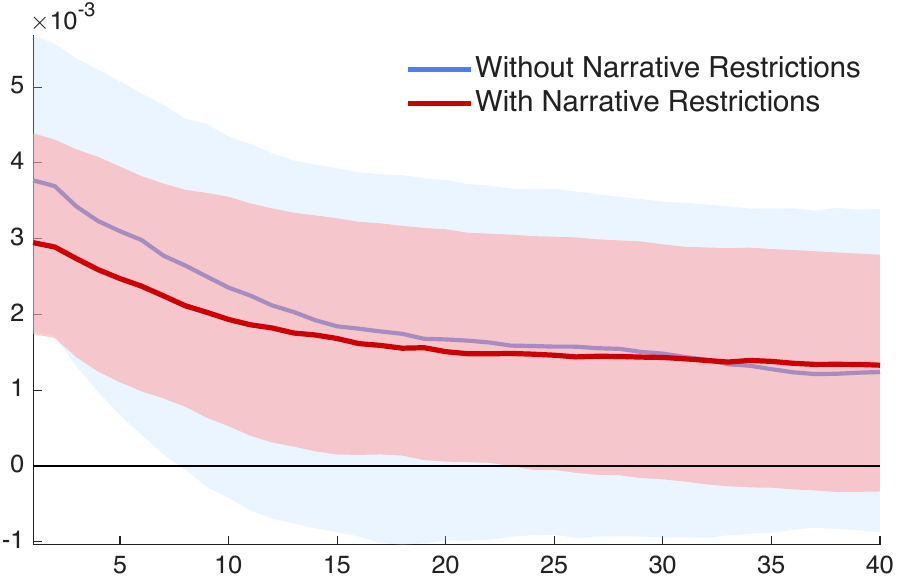}
&
\includegraphics[
    width=0.32\textwidth,
    trim=1mm 1mm 1mm 1mm,
    clip
]{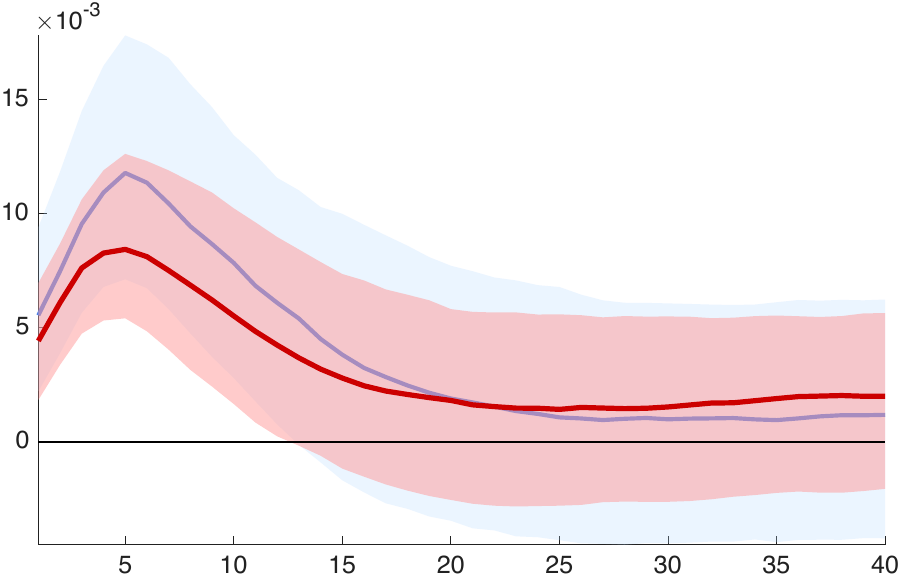}
&
\includegraphics[
    width=0.32\textwidth,
    trim=1mm 1mm 1mm 1mm,
    clip
]{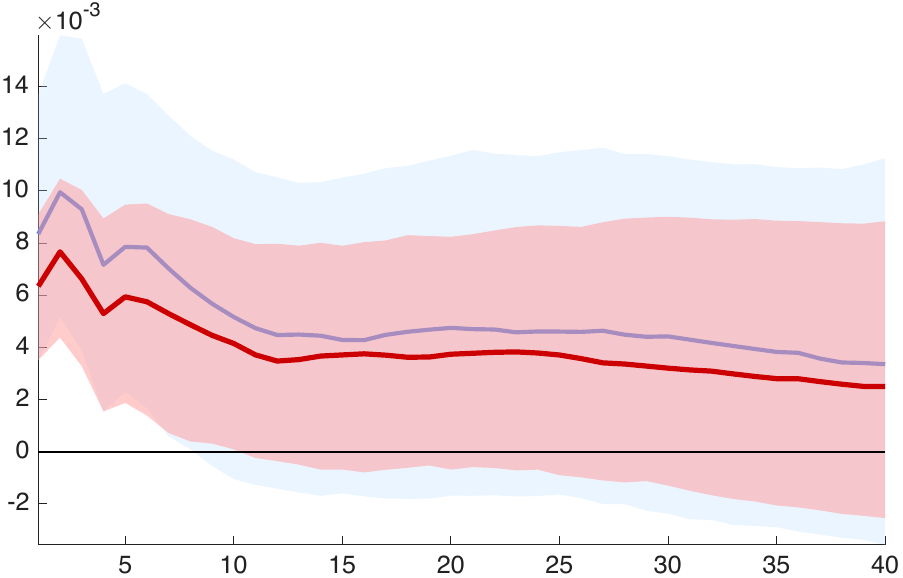}

\\[0.4cm]

GDP Deflator & Unemployment & Disposable Income \\

\includegraphics[
    width=0.32\textwidth,
    trim=1mm 1mm 1mm 1mm,
    clip
]{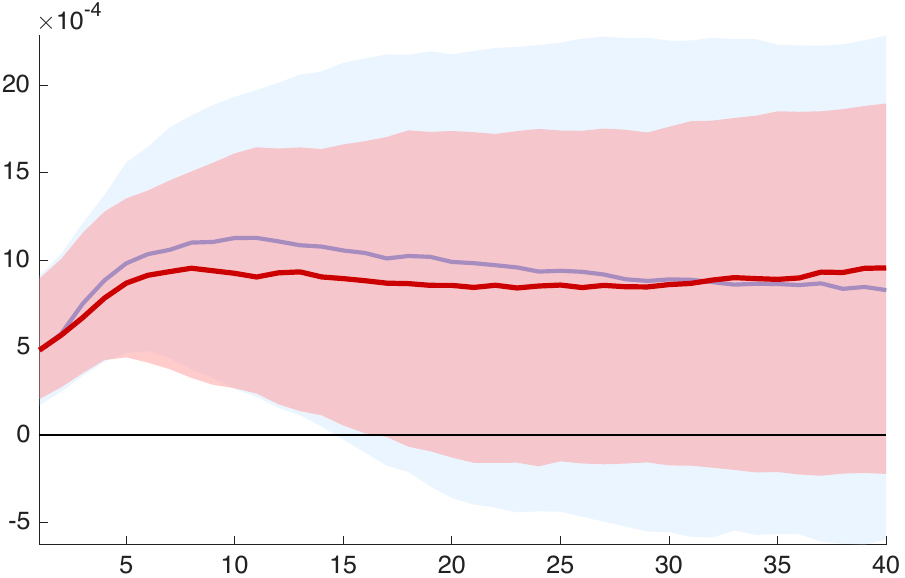}
&
\includegraphics[
    width=0.32\textwidth,
    trim=1mm 1mm 1mm 1mm,
    clip
]{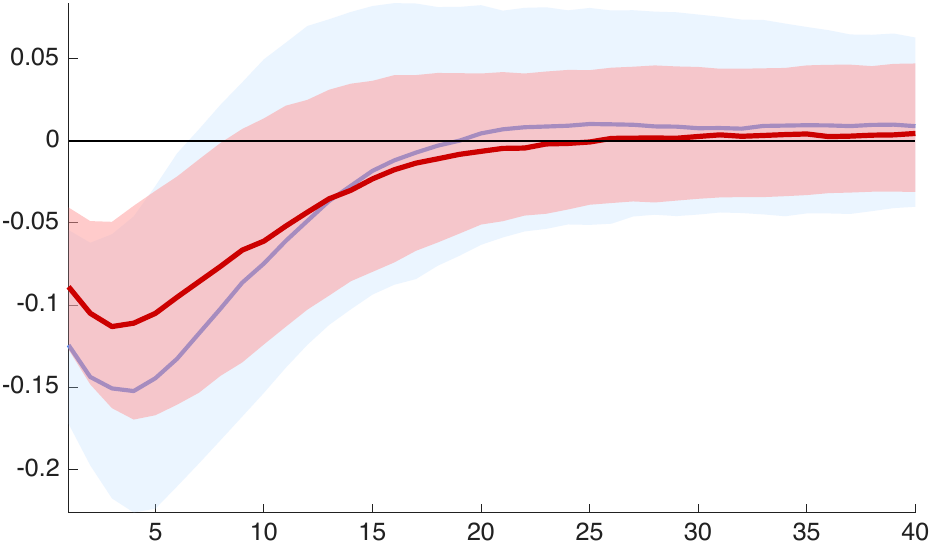}
&
\includegraphics[
    width=0.32\textwidth,
    trim=1mm 1mm 1mm 1mm,
    clip
]{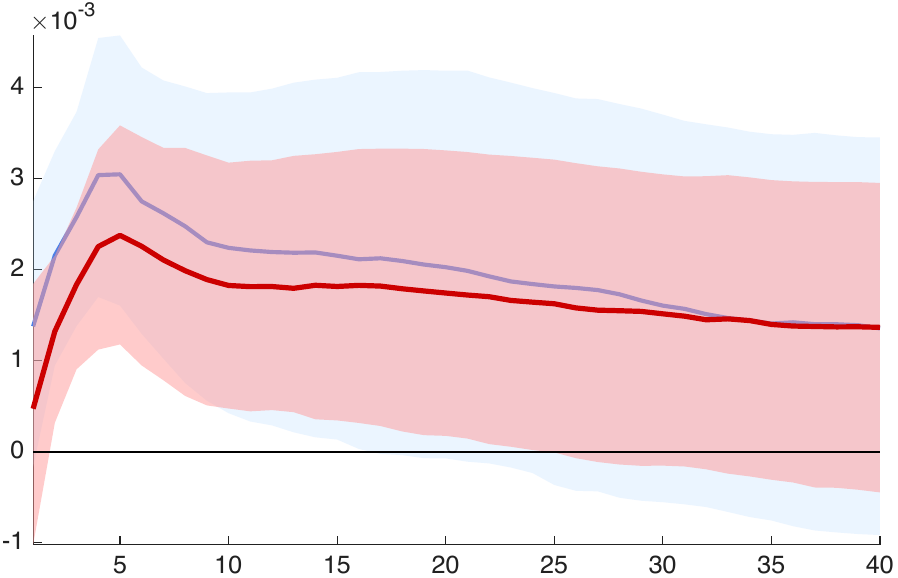}

\end{tabular}
\vspace{-0.25cm} 
\caption{Selected IRFs to Favorable Demand Shock.} 
\label{fig_demand_shock}
\vspace{0.05cm} 
\begin{minipage}{0.98\textwidth}
\footnotesize
\textit{Notes:} Impulse responses of selected economic indicators.
Blue: 68\% credible sets (no narrative restrictions).
Red: median and 68\% bands (with narrative restrictions).
\end{minipage}
\end{figure}

Figure \ref{fig_demand_shock} displays the impulse responses of selected macroeconomic variables to a favorable demand shock. The solid blue line and shaded areas denote the median and 68\% credible bands from the baseline model without narrative restrictions, while the dotted red line and corresponding bands represent the extended specification that augments sign and ranking restrictions with narrative information. This notation is used consistently across all subsequent figures, which report only those shocks for which narrative restrictions are imposed.

Across all horizons, the model with narrative restrictions yields visibly narrower credible bands, reflecting the reduction in the size of the identified set. In addition, the median responses differ between the two specifications, indicating that narrative restrictions affect not only the precision but also the location of the estimated impulse responses. Quantitatively, the narrative-restricted model implies a more muted increase in investment and a smaller decline in unemployment in the first years following the shock.

These differences suggest that, absent narrative information, the identified demand shock may be partially confounded with other structural disturbances, such as investment or financial shocks. Incorporating the narrative restriction for 2006Q1 helps disentangle these effects and leads to more disciplined estimates of both the dynamic responses and their magnitudes.

\begin{figure}[H]
\centering
\setlength{\tabcolsep}{2pt}
\renewcommand{\arraystretch}{1}

\begin{tabular}{ccc}
GDP & Investment & Imports \\

\includegraphics[
    width=0.32\textwidth,
    trim=1mm 1mm 1mm 1mm,
    clip
]{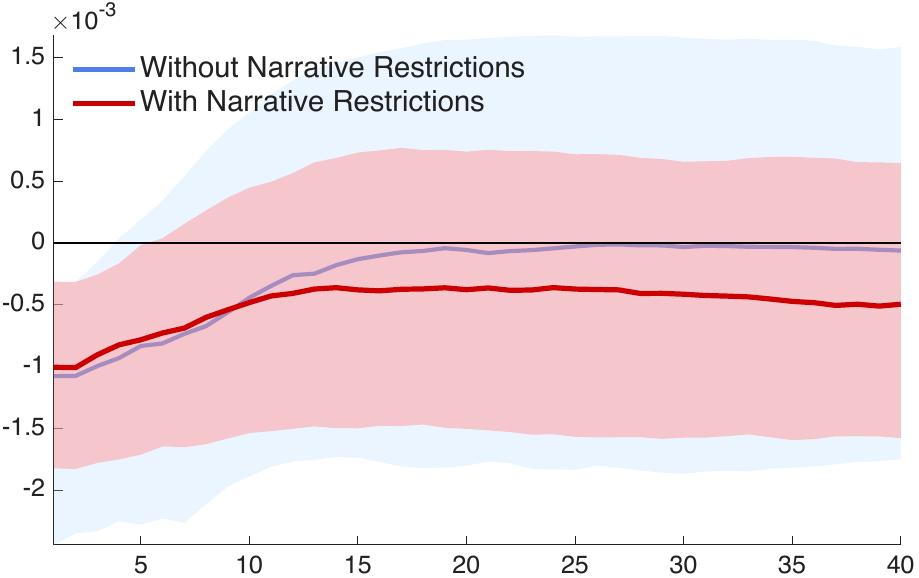}
&
\includegraphics[
    width=0.32\textwidth,
    trim=1mm 1mm 1mm 1mm,
    clip
]{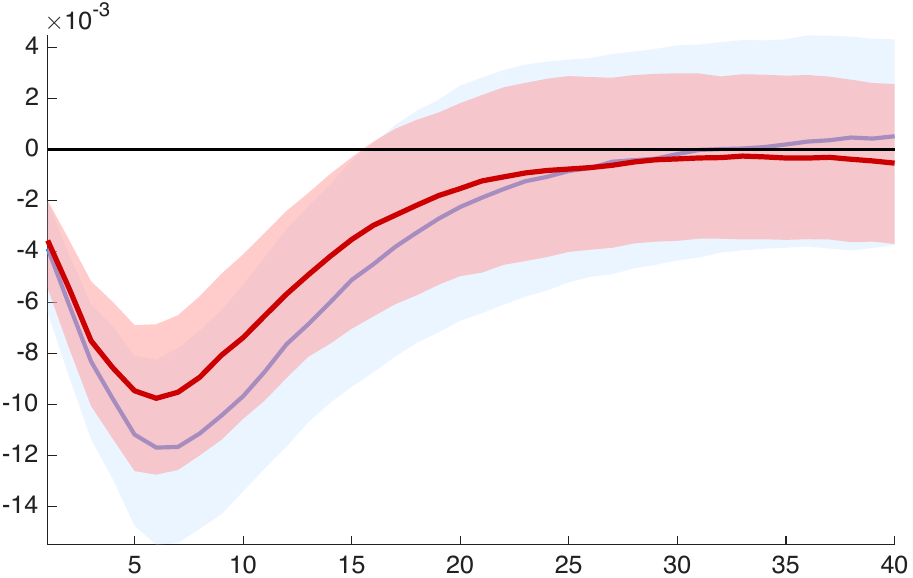}
&
\includegraphics[
    width=0.32\textwidth,
    trim=1mm 1mm 1mm 1mm,
    clip
]{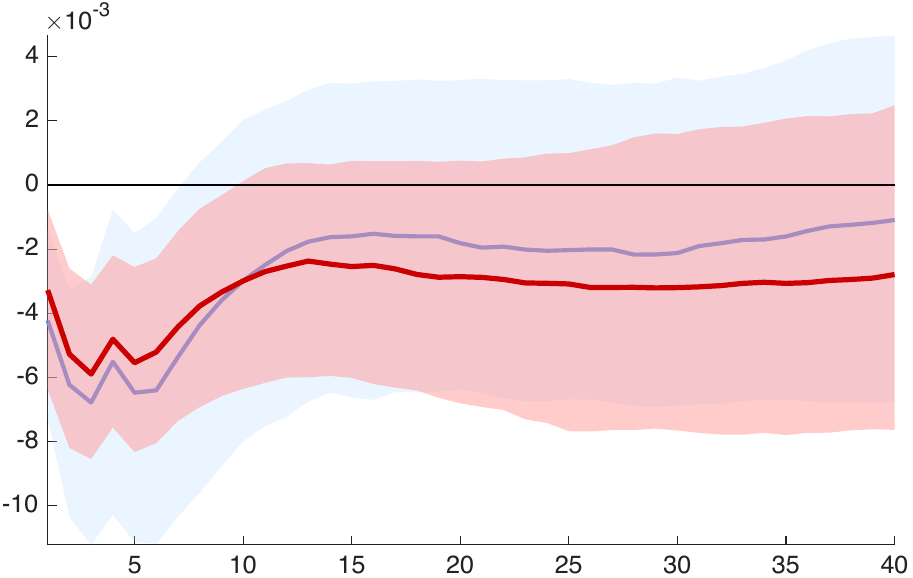}

\\[0.4cm]

GDP Deflator & Unemployment & IP \\

\includegraphics[
    width=0.32\textwidth,
    trim=1mm 1mm 1mm 1mm,
    clip
]{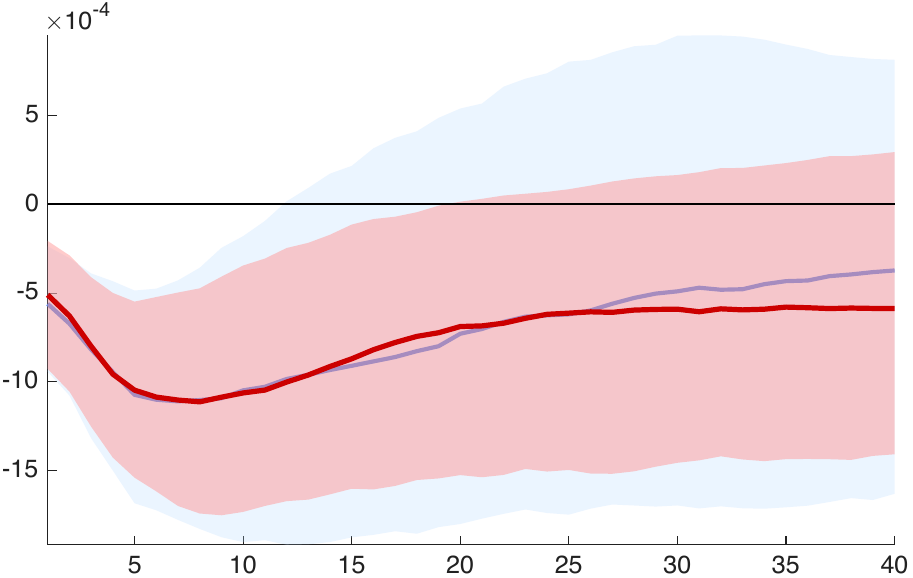}
&
\includegraphics[
    width=0.32\textwidth,
    trim=1mm 1mm 1mm 1mm,
    clip
]{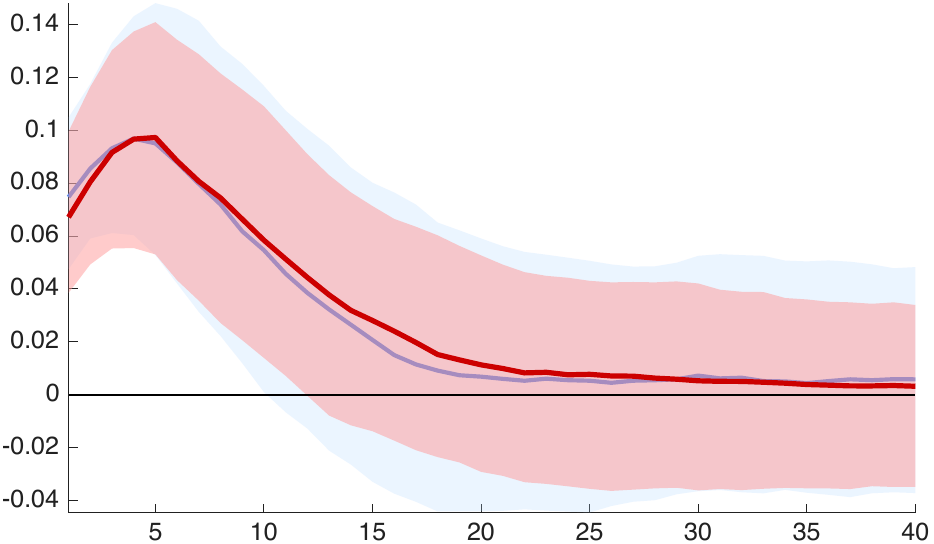}
&
\includegraphics[
    width=0.32\textwidth,
    trim=1mm 1mm 1mm 1mm,
    clip
]{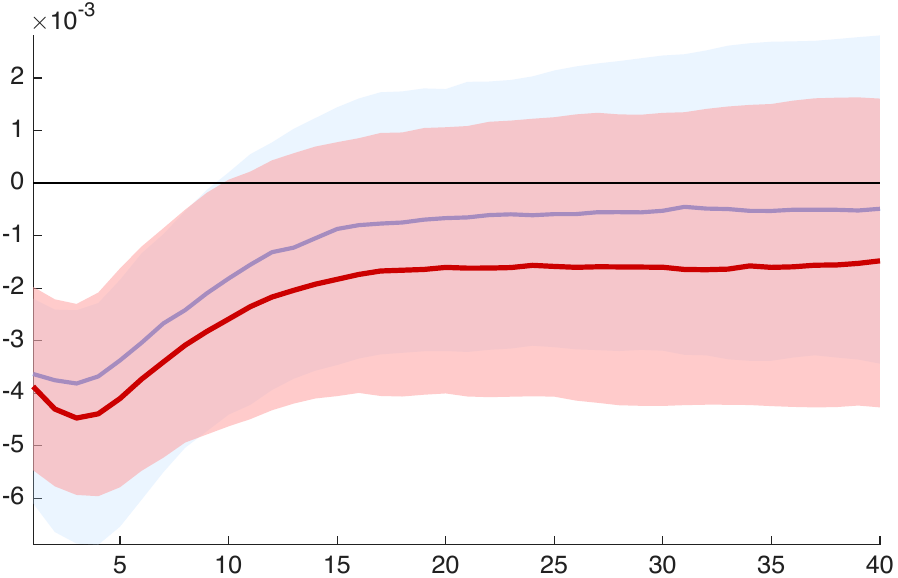}

\end{tabular}
\vspace{-0.25cm} 
\caption{Selected IRFs to Adverse Financial Shock.} 
\label{fig_financial_shock}
\vspace{0.05cm} 
\begin{minipage}{0.98\textwidth}
\footnotesize
\textit{Notes:} Impulse responses of selected economic indicators.
Blue: 68\% credible sets (no narrative restrictions).
Red: median and 68\% bands (with narrative restrictions).
\end{minipage}
\end{figure}

Figure \ref{fig_financial_shock} reports the impulse responses of selected variables to an adverse financial risk shock. As before, the specification that incorporates narrative restrictions yields credible bands that are uniformly no wider—and in several cases noticeably narrower—than those of the baseline model, indicating a sharpened identification.

In terms of point estimates, the median responses differ across specifications. In particular, GDP, imports, and industrial production exhibit a more persistent contraction in the narrative-restricted model, especially at longer horizons. This suggests that incorporating narrative information not only sharpens inference but also shifts the estimated propagation of financial shocks toward more pronounced medium- to long-run effects.

Overall, these results reinforce the view that narrative restrictions help to discipline the identification of financial shocks, thereby reducing ambiguity in both the magnitude and persistence of the associated impulse responses.

\begin{figure}[H]
\centering
\setlength{\tabcolsep}{2pt}
\renewcommand{\arraystretch}{1}

\begin{tabular}{ccc}
GDP & Gov. Spending & GDP Deflator \\

\includegraphics[
    width=0.32\textwidth,
    trim=1mm 1mm 1mm 1mm,
    clip
]{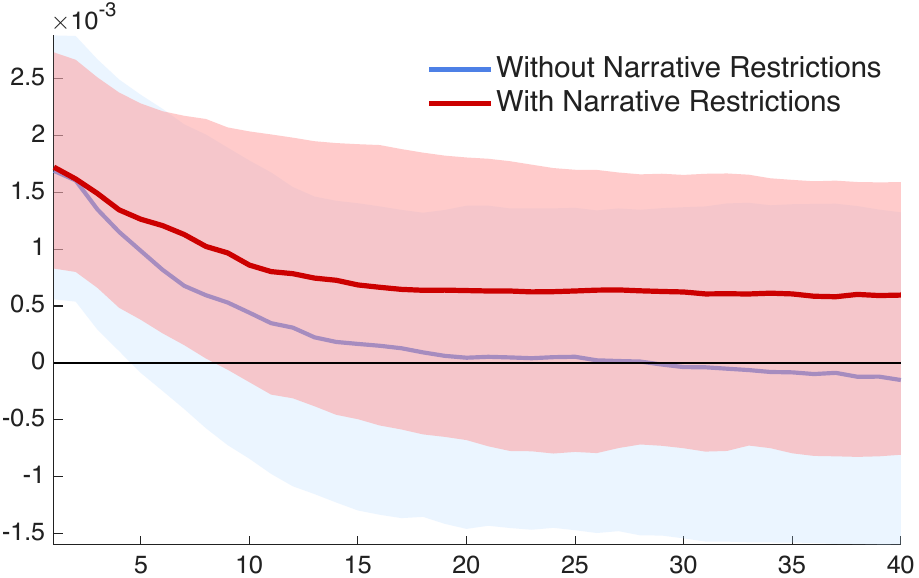}
&
\includegraphics[
    width=0.32\textwidth,
    trim=1mm 1mm 1mm 1mm,
    clip
]{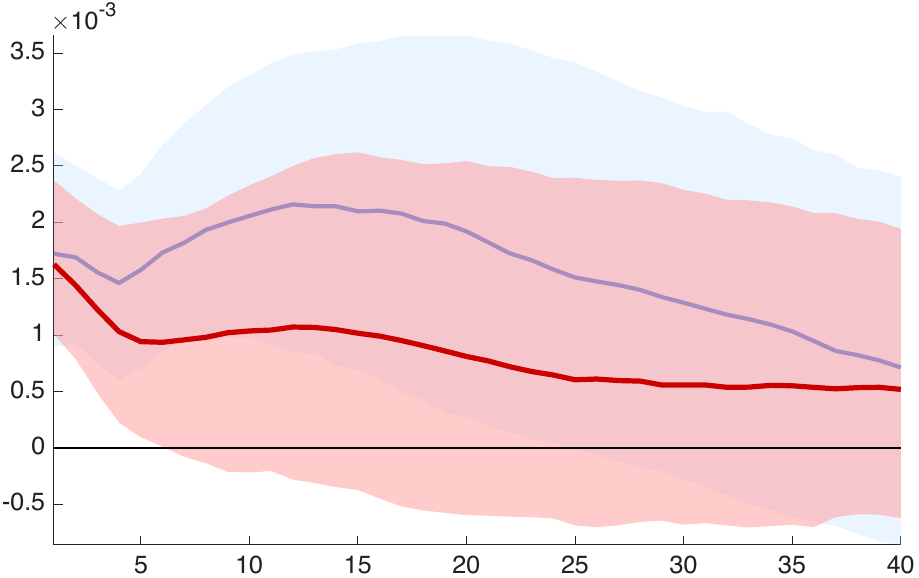}
&
\includegraphics[
    width=0.32\textwidth,
    trim=1mm 1mm 1mm 1mm,
    clip
]{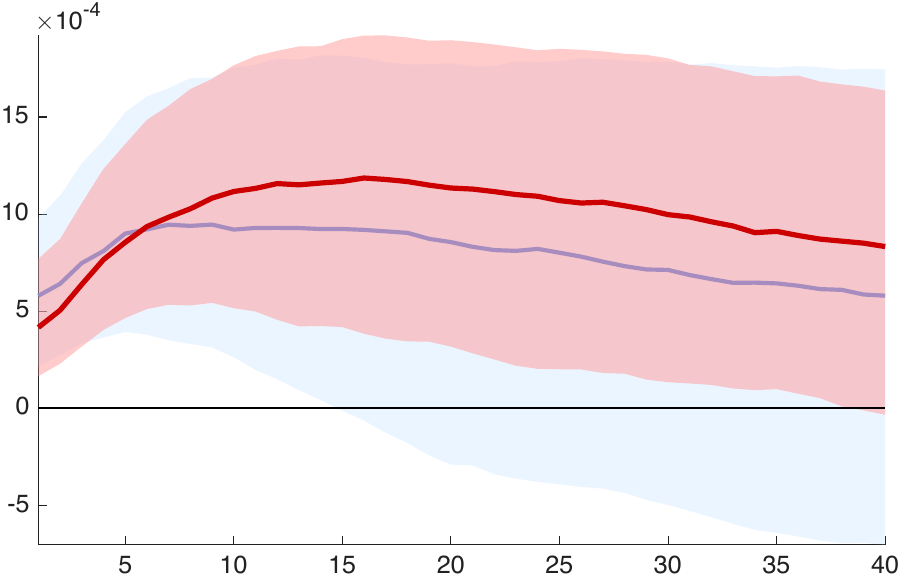}

\\[0.4cm]

CPI & Unemployment & Disposable Income \\

\includegraphics[
    width=0.32\textwidth,
    trim=1mm 1mm 1mm 1mm,
    clip
]{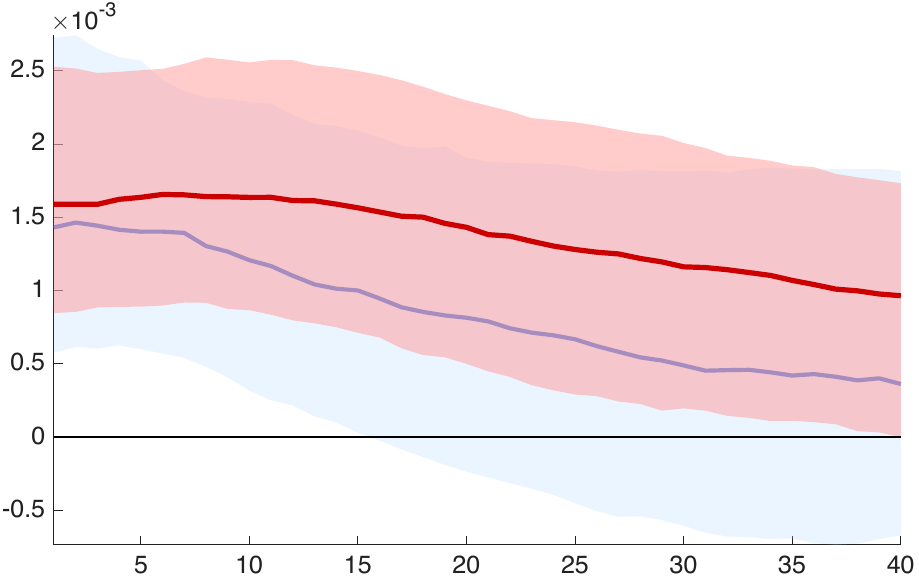}
&
\includegraphics[
    width=0.32\textwidth,
    trim=1mm 1mm 1mm 1mm,
    clip
]{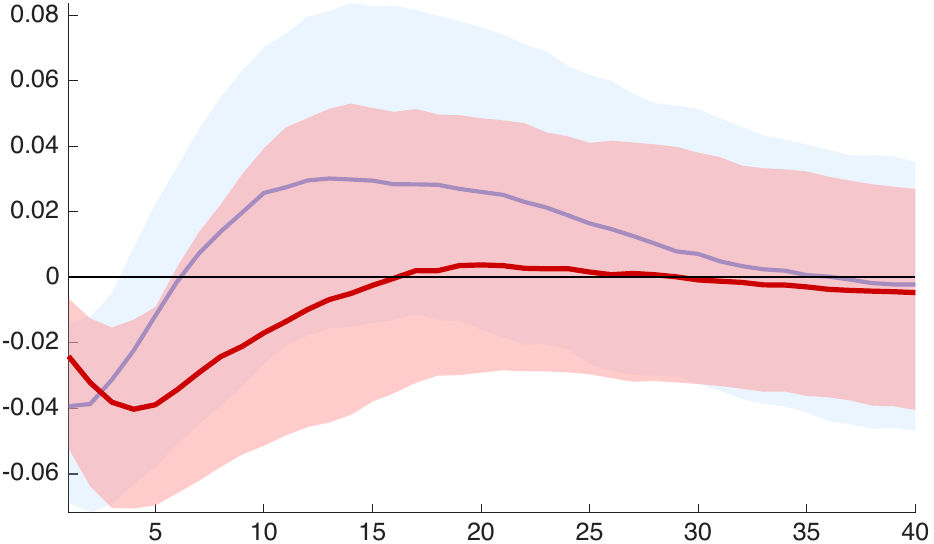}
&
\includegraphics[
    width=0.32\textwidth,
    trim=1mm 1mm 1mm 1mm,
    clip
]{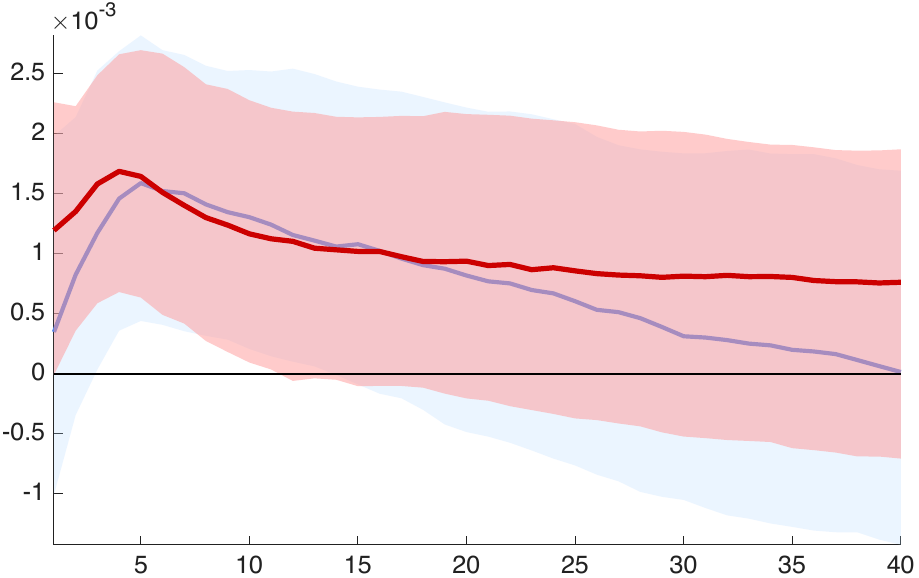}

\end{tabular}
\vspace{-0.25cm} 
\caption{Selected IRFs to Expansionary Government Spending Shock.} 
\label{fig_government_shock}
\vspace{0.05cm} 
\begin{minipage}{0.98\textwidth}
\footnotesize
\textit{Notes:} Impulse responses of selected economic indicators.
Blue: 68\% credible sets (no narrative restrictions).
Red: median and 68\% bands (with narrative restrictions).
\end{minipage}
\end{figure}

Figure \ref{fig_government_shock} presents selected impulse responses to an expansionary government spending shock. Consistent with the previous results, the specification with narrative restrictions yields credible bands that are at least as narrow as those of the baseline model, again reflecting an improvement in identification.

Regarding the median responses, the effects differ across specifications. Government spending itself exhibits a more short-lived increase in the narrative-restricted model compared to the baseline. At the same time, GDP, unemployment, and disposable income display stronger responses, indicating a more pronounced transmission of the fiscal shock to real economic activity. Most notably, price dynamics—as captured by the GDP deflator and CPI—are both more pronounced and more persistent in the extended model.

These findings suggest that imposing narrative restrictions for the identified episodes in 1980Q1 and 2011Q3 helps to sharpen the identification of government spending shocks, leading to a clearer and more consistent characterization of their macroeconomic effects.

\begin{figure}[H]
\centering
\setlength{\tabcolsep}{2pt}
\renewcommand{\arraystretch}{1}

\begin{tabular}{ccc}
GDP & Investment & CPI \\

\includegraphics[
    width=0.32\textwidth,
    trim=1mm 1mm 1mm 1mm,
    clip
]{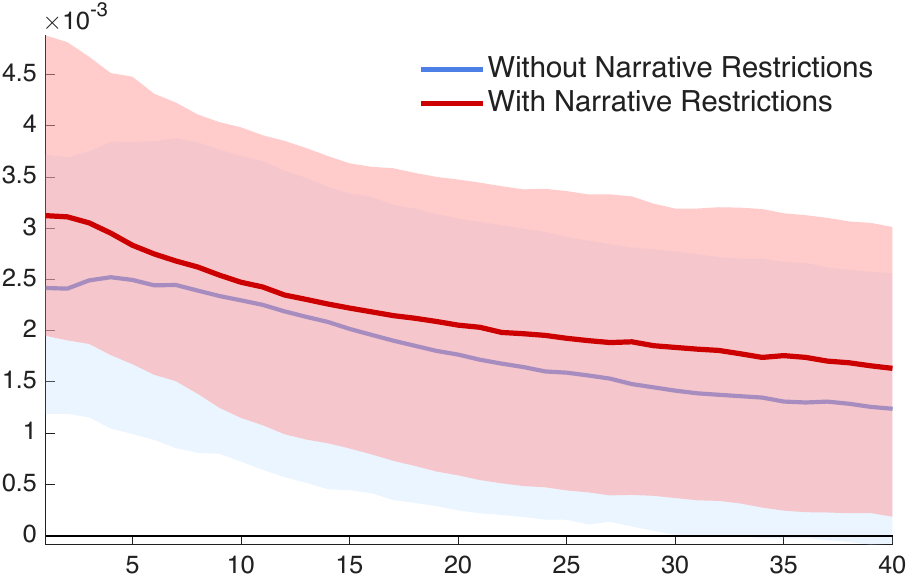}
&
\includegraphics[
    width=0.32\textwidth,
    trim=1mm 1mm 1mm 1mm,
    clip
]{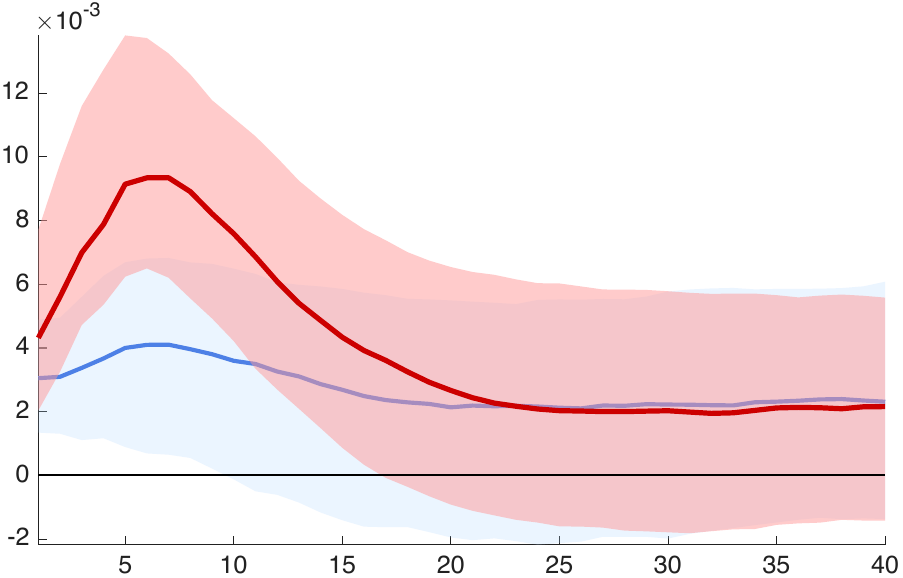}
&
\includegraphics[
    width=0.32\textwidth,
    trim=1mm 1mm 1mm 1mm,
    clip
]{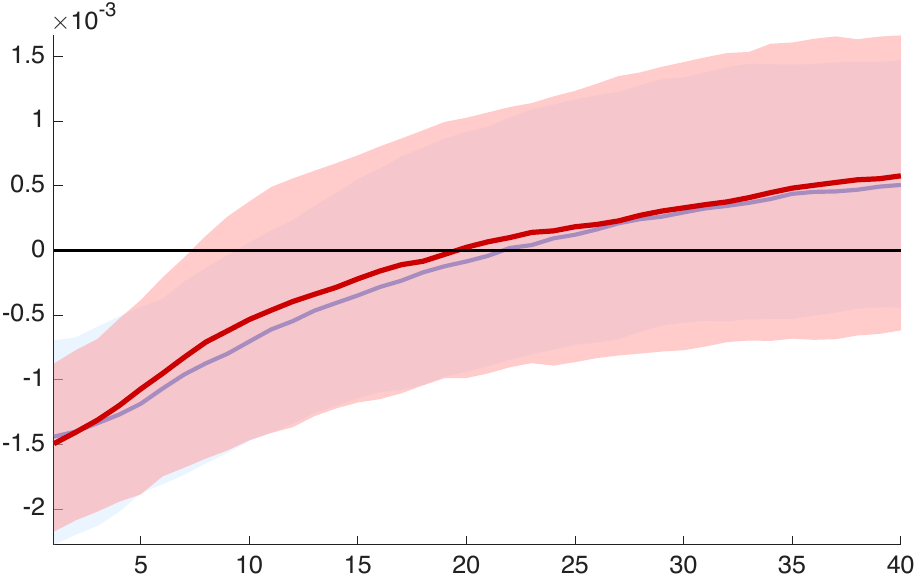}

\\[0.4cm]

Unemployment & Disposable Income & S\&P 500 \\

\includegraphics[
    width=0.32\textwidth,
    trim=1mm 1mm 1mm 1mm,
    clip
]{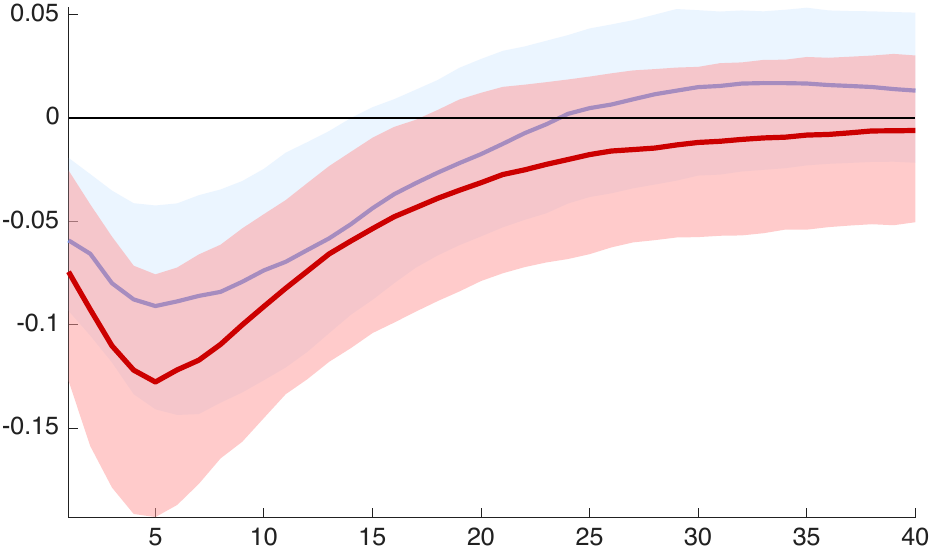}
&
\includegraphics[
    width=0.32\textwidth,
    trim=1mm 1mm 1mm 1mm,
    clip
]{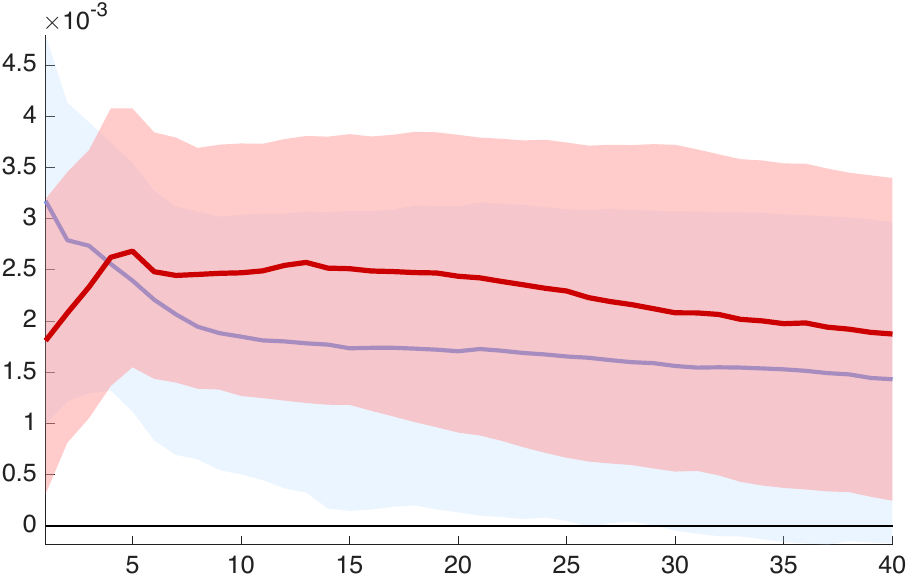}
&
\includegraphics[
    width=0.32\textwidth,
    trim=1mm 1mm 1mm 1mm,
    clip
]{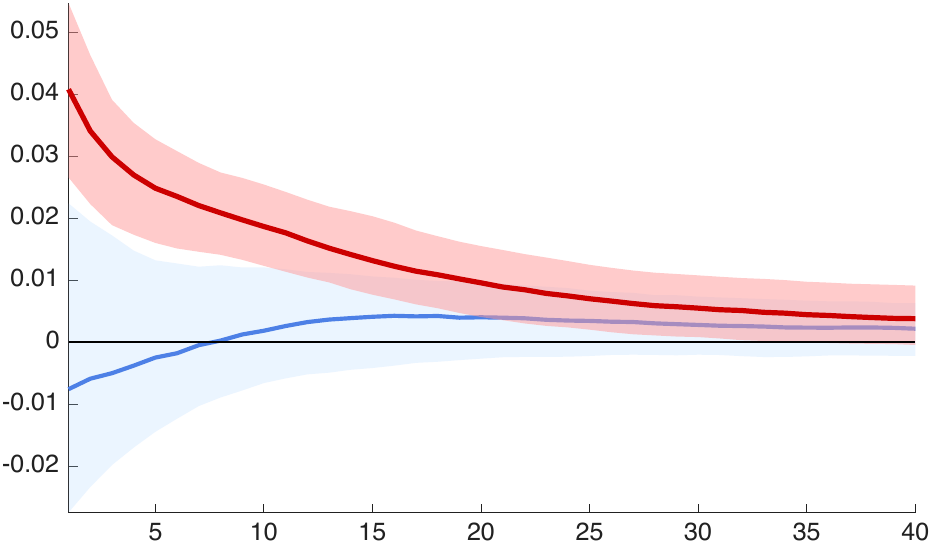}

\end{tabular}
\vspace{-0.25cm} 
\caption{Selected IRFs to Favorable Oil Supply Shock.} 
\label{fig_oil_shock}
\vspace{0.05cm} 
\begin{minipage}{0.98\textwidth}
\footnotesize
\textit{Notes:} Impulse responses of selected economic indicators.
Blue: 68\% credible sets (no narrative restrictions).
Red: median and 68\% bands (with narrative restrictions).
\end{minipage}
\end{figure}

Figure \ref{fig_oil_shock} presents selected impulse responses to an oil supply shock. In contrast to the previous cases, the effects of narrative restrictions are somewhat more mixed: while some responses exhibit narrower credible bands, others are of comparable width or slightly wider than in the baseline model. This suggests that the tightening in identification is less uniform for the oil supply shock.

Despite this, incorporating narrative restrictions for the episodes in 1990Q3, 2002Q4, 2003Q1, and 2011Q1 leads to a clearer and more coherent pattern of responses. In particular, investment, unemployment, and disposable income display more pronounced reactions in the narrative-restricted specification. Most notably, the response of stock prices switches from being negative and statistically indistinguishable from zero in the baseline model to positive and statistically significant when narrative restrictions are imposed.

Overall, these results indicate that, even when the impact on credible band width is less systematic, narrative restrictions can materially improve the interpretability of structural shocks by sharpening both the sign and the statistical significance of key responses.

\begin{figure}[H]
\centering
\setlength{\tabcolsep}{2pt}
\renewcommand{\arraystretch}{1}

\begin{tabular}{ccc}
Demand Shock in 2006Q1 & Fin. Shock in 2007Q3 & Gov. Shock in 1980Q1 \\

\includegraphics[
    width=0.32\textwidth,
    trim=1mm 1mm 1mm 1mm,
    clip
]{"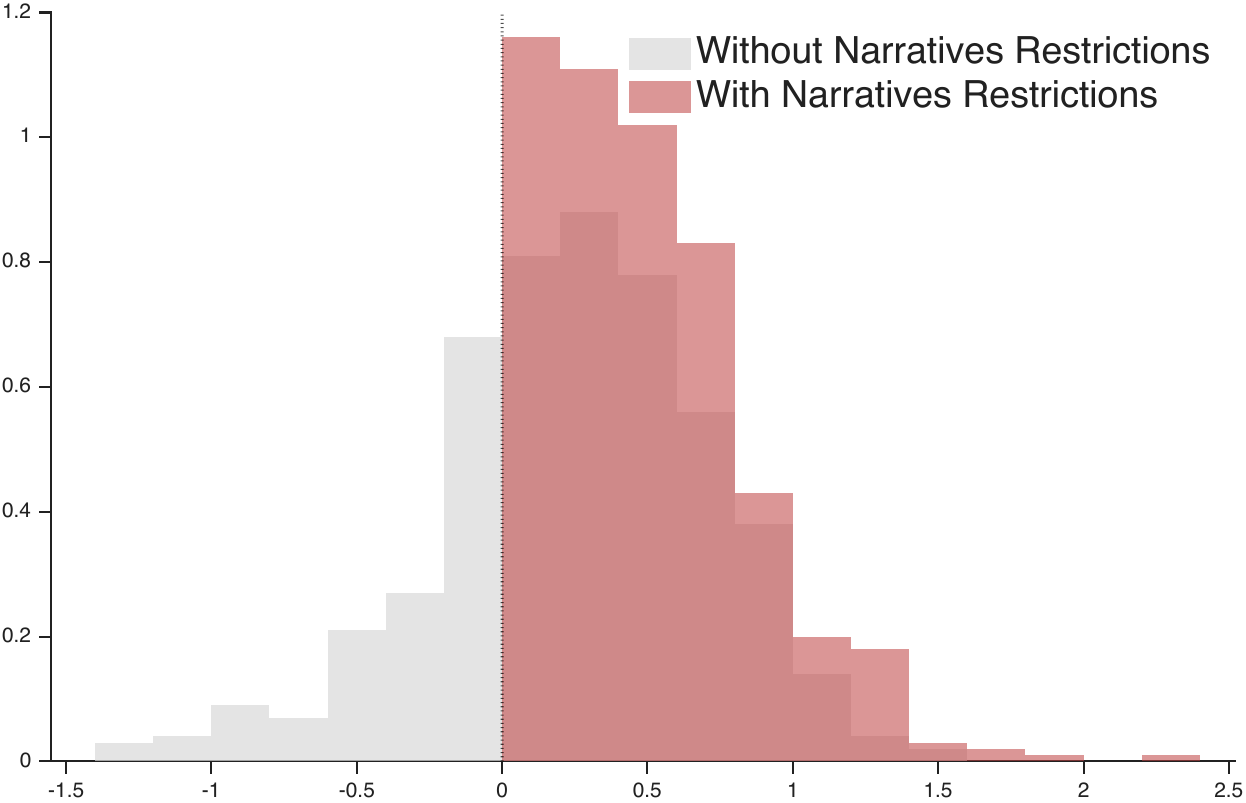"}
&
\includegraphics[
    width=0.32\textwidth,
    trim=1mm 1mm 1mm 1mm,
    clip
]{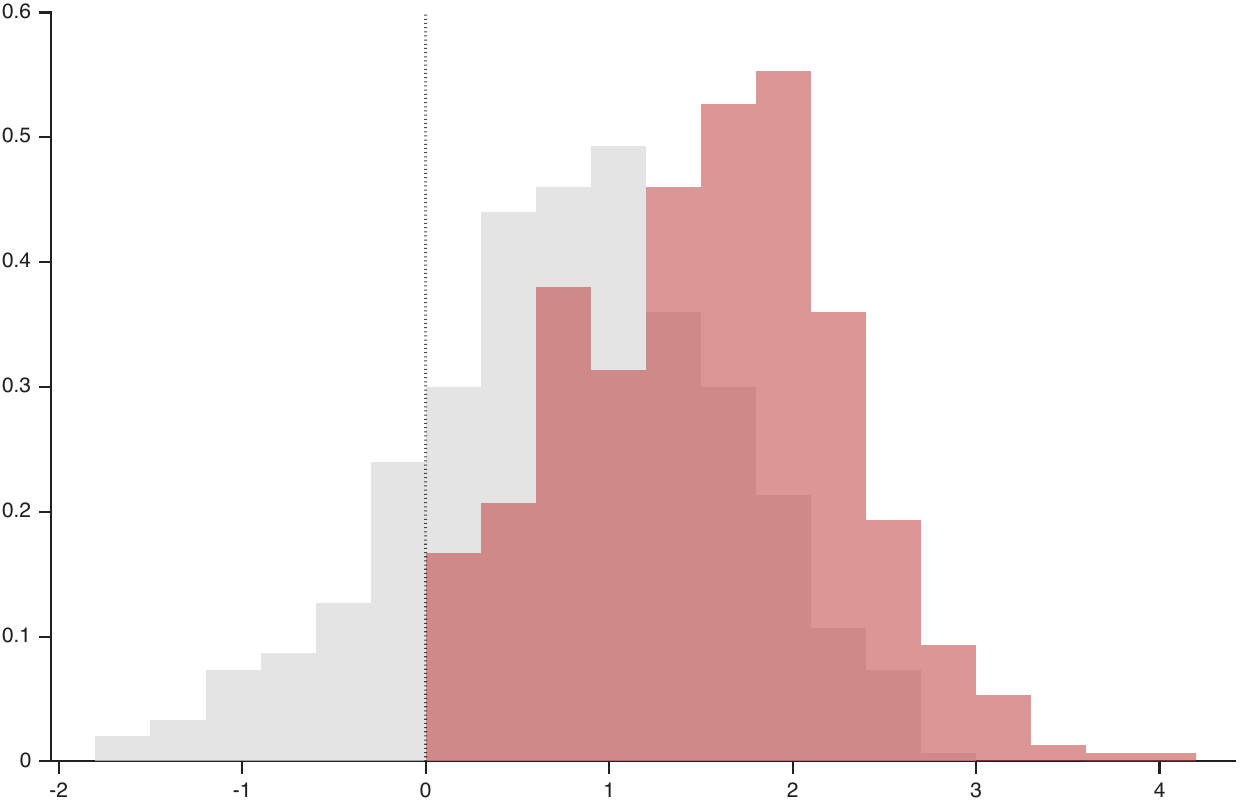}
&
\includegraphics[
    width=0.32\textwidth,
    trim=1mm 1mm 1mm 1mm,
    clip
]{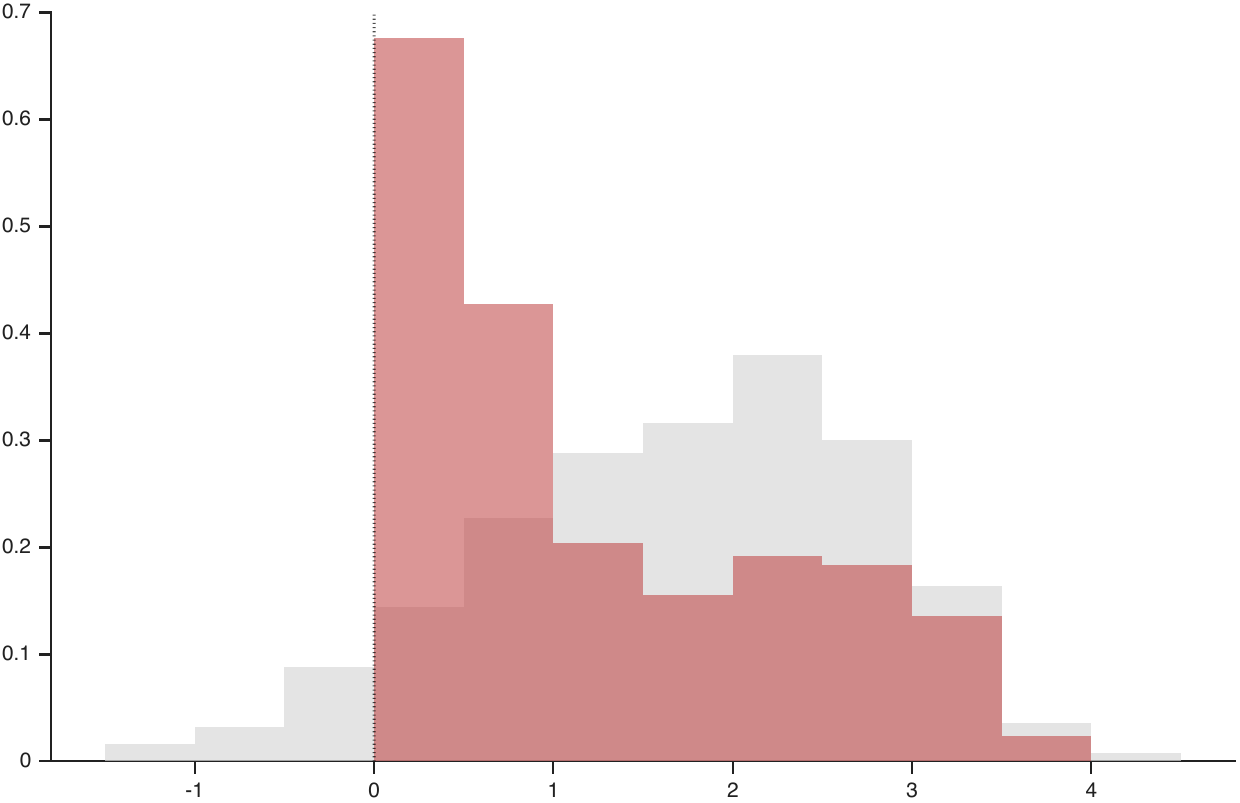}

\\[0.4cm]

Gov. Shock in 2001Q3 & Oil Shock in 1990Q3 &Oil Shock in 2002Q4 \\

\includegraphics[
    width=0.32\textwidth,
    trim=1mm 1mm 1mm 1mm,
    clip
]{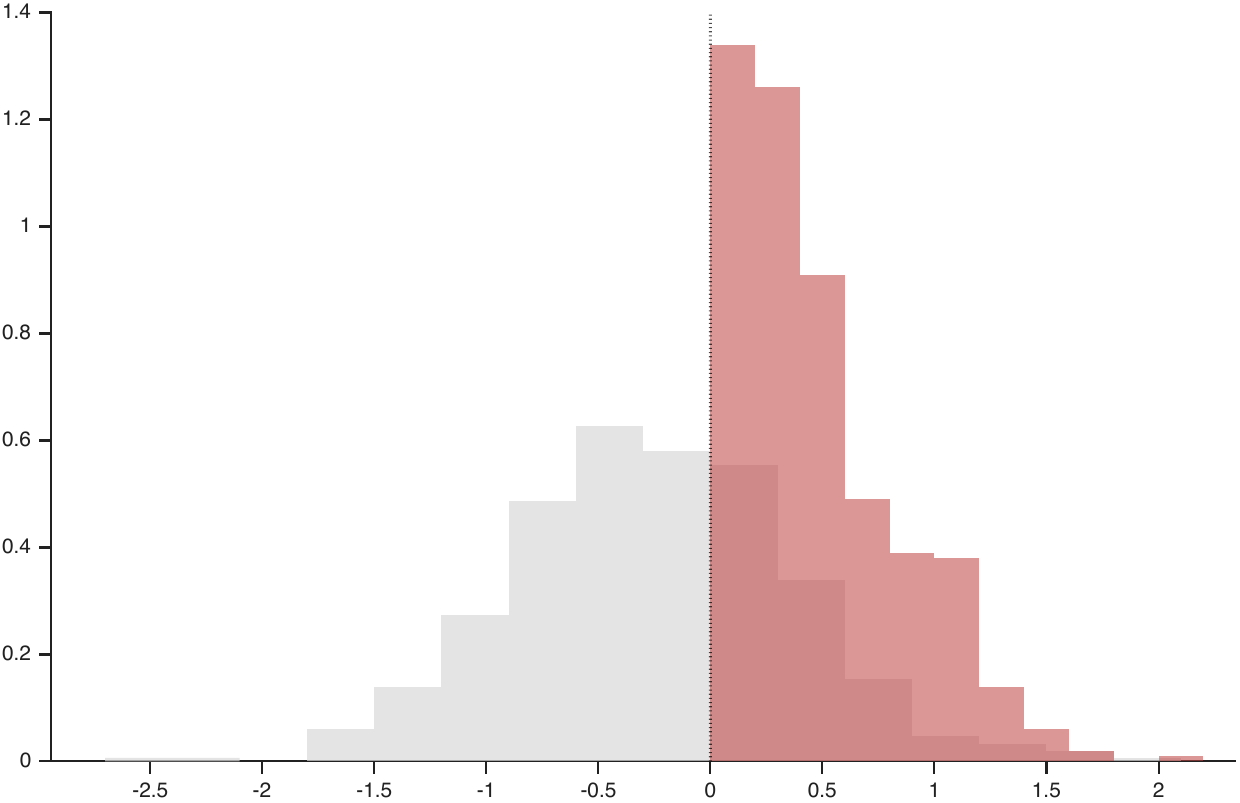}
&
\includegraphics[
    width=0.32\textwidth,
    trim=1mm 1mm 1mm 1mm,
    clip
]{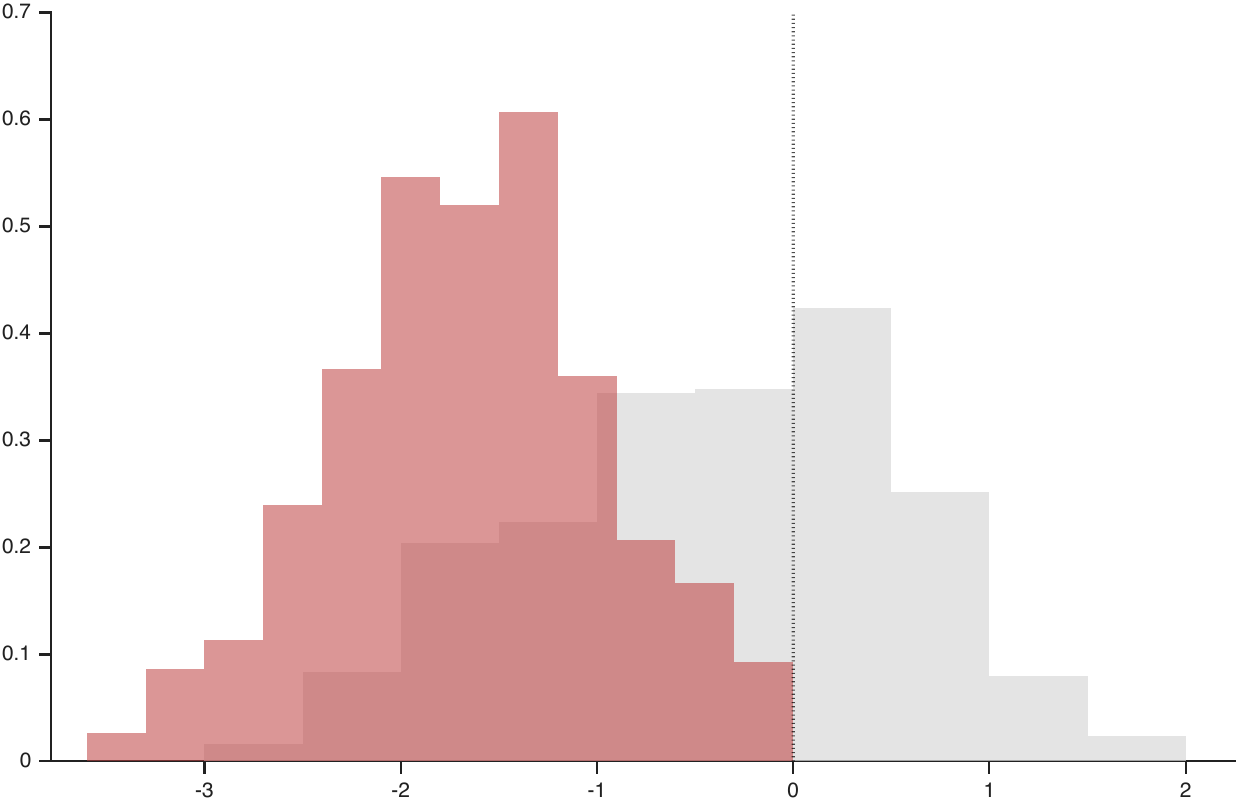}
&
\includegraphics[
    width=0.32\textwidth,
    trim=1mm 1mm 1mm 1mm,
    clip
]{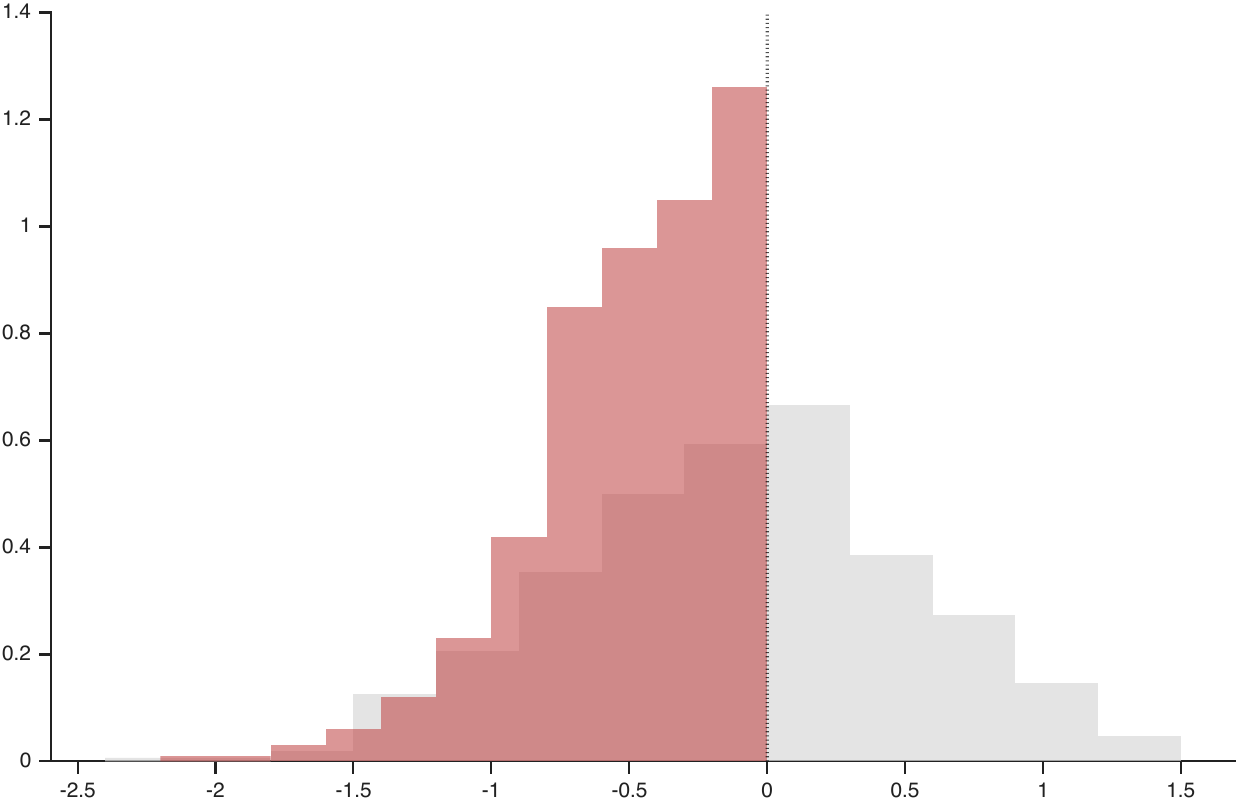}

\end{tabular}
\vspace{-0.25cm} 
\caption{Selected Posterior Distributions of Selected Shocks.} 
\label{fig_shock_distribution}
\vspace{0.05cm} 
\begin{minipage}{0.98\textwidth}
\footnotesize
\textit{Notes:} The light gray histograms show the posterior distributions of the selected structural shocks in a given quarter for the baseline model. The light red histograms show the corresponding posterior distributions for the model with narrative restrictions.
\end{minipage}
\end{figure}

In Figure \ref{fig_shock_distribution}, we present selected posterior distributions of the structural shocks for which narrative restrictions are imposed, evaluated at the specific quarters. For the periods shown, the introduction of narrative restrictions both narrows the support of the posterior distributions and, importantly, shifts it toward positive (or negative) values. With the exception of the expansionary government shock in 1980Q1, the narrative restrictions induce a substantial shift in the posterior mass toward the imposed sign. This highlights both the effectiveness of the proposed algorithm and the importance of a careful and economically sound selection of narrative restrictions.

As a robustness check, we re-estimate the model using alternative sets of narrative restrictions. In Appendix \ref{app_add_emp_results}, we report selected impulse responses for demand, financial, government spending, and oil supply shocks when augmenting the extended specification with three additional narrative restrictions on oil supply shocks. Specifically, we impose negative oil supply shocks in 1978Q4 (Iranian Revolution), as well as in 1980Q3 and 1980Q4 (Iran–Iraq War) (see Table \ref{tab:shock-quarter-restrictions_alternative}). Figures \ref{fig2_demand_shock}–\ref{fig2_oil_shock} present the corresponding impulse responses for the same set of variables as in the above analysis.

Across these alternative specifications, the main qualitative findings remain unchanged, indicating that the results are robust to the inclusion of additional historically motivated narrative restrictions. We further consider alternative combinations of narrative restrictions, as well as a different lag length for the VAR. These additional results are available upon request.

\section{Conclusion}

This paper develops a high-dimensional structural VAR framework that accommodates a large number of linear inequality restrictions on both impact impulse responses and structural shocks. By combining a factor structure in the error terms with a computationally efficient sampling algorithm that avoids rejection-based procedures, our approach overcomes a key limitation of existing methods: their limited scalability when the dimension of the system and the number of restrictions increase. This makes it feasible to jointly impose rich sets of impact, and shock restrictions in large-scale systems. More broadly, our framework extends recent advances in large sign-restricted VARs by allowing narrative restrictions to be imposed directly through constraints on structural shocks. This additional flexibility sharpens identification and enhances the economic interpretability of the identified disturbances.

Overall, the proposed framework substantially broadens the scope for credible, flexible, and computationally tractable structural analysis in high-dimensional VAR environments. More generally, it provides a practical foundation for combining rich information sets with economically motivated identification restrictions, opening new possibilities for structural inference in large macroeconomic systems.

In our empirical application, we illustrate the usefulness of this framework in a large-scale U.S. VAR with 39 variables and ten structural shocks. We show that incorporating narrative restrictions alongside sign and ranking restrictions reduces the size of the identified set and yields more precisely estimated impulse responses. Across a range of shocks, this leads to narrower credible bands and, in several cases, materially different median responses, highlighting the empirical relevance of our approach for applied macroeconomic analysis.

\newpage
\small
\setstretch{0.01} 
\addcontentsline{toc}{section}{References}
\bibliography{literature.bib}

\begin{thebibliography}{}

\bibitem[Altavilla et~al., 2019]{altavilla2019loan}
Altavilla, C., Paries, M.~D., and Nicoletti, G. (2019).
\newblock Loan supply, credit markets and the euro area financial crisis.
\newblock {\em Journal of Banking \& Finance}, 109:105658.

\bibitem[Anderson and Rubin, 1956]{anderson1956statistical}
Anderson, T. and Rubin, H. (1956).
\newblock Statistical inference in factor analysis.
\newblock In {\em Proceedings of the Berkeley Symposium on Mathematical Statistics and Probability}, volume~5, pages 111--150. University of California Press.

\bibitem[Antolin-Diaz et~al., 2021]{antolin2021structural}
Antolin-Diaz, J., Petrella, I., and Rubio-Ram{\'\i}rez, J.~F. (2021).
\newblock Structural scenario analysis with svars.
\newblock {\em Journal of Monetary Economics}, 117:798--815.

\bibitem[Antol{\'\i}n-D{\'\i}az and Rubio-Ram{\'\i}rez, 2018]{antolin2018narrative}
Antol{\'\i}n-D{\'\i}az, J. and Rubio-Ram{\'\i}rez, J.~F. (2018).
\newblock Narrative sign restrictions for svars.
\newblock {\em American Economic Review}, 108(10):2802--2829.

\bibitem[Arias et~al., 2025]{arias2025large}
Arias, J.~E., Rubio-Ramirez, J.~F., and Shin, M. (2025).
\newblock Large svars.

\bibitem[Ascari et~al., 2023]{ascari2023endogenous}
Ascari, G., Fasani, S., Grazzini, J., and Rossi, L. (2023).
\newblock Endogenous uncertainty and the macroeconomic impact of shocks to inflation expectations.
\newblock {\em Journal of Monetary Economics}, 140:S48--S63.

\bibitem[Badinger and Schiman, 2023a]{badinger2023measuring}
Badinger, H. and Schiman, S. (2023a).
\newblock Measuring monetary policy in the euro area using svars with residual restrictions.
\newblock {\em American Economic Journal: Macroeconomics}, 15(2):279--305.

\bibitem[Badinger and Schiman, 2023b]{Badinger2023}
Badinger, H. and Schiman, S. (2023b).
\newblock Measuring monetary policy in the euro area using svars with residual restrictions.
\newblock {\em American Economic Journal: Macroeconomics}, 15(2):279--305.

\bibitem[Bai, 2003]{bai2003}
Bai, J. (2003).
\newblock Inferential theory for factor models of large dimensions.
\newblock {\em Econometrica}, 71(1):135--171.

\bibitem[Baker et~al., 2016]{baker2016measuring}
Baker, S.~R., Bloom, N., and Davis, S.~J. (2016).
\newblock Measuring economic policy uncertainty.
\newblock {\em The quarterly journal of economics}, 131(4):1593--1636.

\bibitem[Banbura et~al., 2023]{banbura2023drives}
Banbura, M., Bobeica, E., and Hern{\'a}ndez, C.~M. (2023).
\newblock What drives core inflation? {T}he role of supply shocks.

\bibitem[Ba{\'n}bura et~al., 2010]{banbura2010large}
Ba{\'n}bura, M., Giannone, D., and Reichlin, L. (2010).
\newblock Large bayesian vector auto regressions.
\newblock {\em Journal of applied Econometrics}, 25(1):71--92.

\bibitem[Berger et~al., 2022]{berger2022unified}
Berger, T., Richter, J., and Wong, B. (2022).
\newblock A unified approach for jointly estimating the business and financial cycle, and the role of financial factors.
\newblock {\em Journal of Economic Dynamics and Control}, 136:104315.

\bibitem[Berthold, 2023]{berthold2023macroeconomic}
Berthold, B. (2023).
\newblock The macroeconomic effects of uncertainty and risk aversion shocks.
\newblock {\em European Economic Review}, 154:104442.

\bibitem[Bobeica et~al., 2025]{bobeica2025beware}
Bobeica, E., Holton, S., Huber, F., and Hern{\'a}ndez, C.~M. (2025).
\newblock Beware of large shocks! a non-parametric structural inflation model.

\bibitem[Boer et~al., 2024]{boer2024energy}
Boer, L., Pescatori, A., and Stuermer, M. (2024).
\newblock Energy transition metals: bottleneck for net-zero emissions?
\newblock {\em Journal of the European Economic Association}, 22(1):200--229.

\bibitem[Botev, 2017]{botev2017normal}
Botev, Z.~I. (2017).
\newblock The normal law under linear restrictions: simulation and estimation via minimax tilting.
\newblock {\em Journal of the Royal Statistical Society Series B: Statistical Methodology}, 79(1):125--148.

\bibitem[Brianti, 2025]{brianti2025financial}
Brianti, M. (2025).
\newblock Financial shocks, uncertainty shocks, and corporate liquidity.
\newblock {\em Journal of Applied Econometrics}, 40(7):814--828.

\bibitem[Caggiano and Castelnuovo, 2023]{caggiano2023global}
Caggiano, G. and Castelnuovo, E. (2023).
\newblock Global financial uncertainty.
\newblock {\em Journal of Applied Econometrics}, 38(3):432--449.

\bibitem[Caggiano et~al., 2021]{caggiano2021financial}
Caggiano, G., Castelnuovo, E., Delrio, S., and Kima, R. (2021).
\newblock Financial uncertainty and real activity: The good, the bad, and the ugly.
\newblock {\em European Economic Review}, 136:103750.

\bibitem[Canova and Paustian, 2011]{canova2011business}
Canova, F. and Paustian, M. (2011).
\newblock Business cycle measurement with some theory.
\newblock {\em Journal of Monetary Economics}, 58(4):345--361.

\bibitem[Carriero et~al., 2009]{carriero2009forecasting}
Carriero, A., Kapetanios, G., and Marcellino, M. (2009).
\newblock Forecasting exchange rates with a large bayesian var.
\newblock {\em International Journal of Forecasting}, 25(2):400--417.

\bibitem[Carvalho et~al., 2010]{carvalho2010horseshoe}
Carvalho, C.~M., Polson, N.~G., and Scott, J.~G. (2010).
\newblock The horseshoe estimator for sparse signals.
\newblock {\em Biometrika}, 97(2):465--480.

\bibitem[Chan et~al., 2022]{chan2022large}
Chan, J., Eisenstat, E., and Yu, X. (2022).
\newblock Large bayesian vars with factor stochastic volatility: Identification, order invariance and structural analysis.
\newblock {\em arXiv preprint arXiv:2207.03988}.

\bibitem[Chan et~al., 2019]{chan2019bayesian}
Chan, J., Koop, G., Poirier, D.~J., and Tobias, J.~L. (2019).
\newblock {\em Bayesian econometric methods}, volume~7.
\newblock Cambridge University Press.

\bibitem[Chan et~al., 2024]{chan2024large}
Chan, J.~C., Koop, G., and Yu, X. (2024).
\newblock Large order-invariant bayesian vars with stochastic volatility.
\newblock {\em Journal of Business \& Economic Statistics}, 42(2):825--837.

\bibitem[Chan et~al., 2025]{chan2025rankingrestrictions}
Chan, J.~C., Matthes, C., and Yu, X. (2025).
\newblock Large structural vars with multiple sign and ranking restrictions.
\newblock {\em working paper}.

\bibitem[Chan and Qi, 2025]{chan2025large}
Chan, J.~C. and Qi, Y. (2025).
\newblock Large bayesian matrix autoregressions.
\newblock {\em Journal of Econometrics}, page 105955.

\bibitem[Cheng and Yang, 2020]{cheng2020revisiting}
Cheng, K. and Yang, Y. (2020).
\newblock Revisiting the effects of monetary policy shocks: Evidence from svar with narrative sign restrictions.
\newblock {\em Economics Letters}, 196:109598.

\bibitem[Clark et~al., 2025]{clark2025nonparametric}
Clark, T., Huber, F., and Koop, G. (2025).
\newblock A nonparametric approach to augmenting a bayesian var with nonlinear factors.
\newblock {\em arXiv preprint arXiv:2508.13972}.

\bibitem[Conti et~al., 2023]{conti2023bank}
Conti, A.~M., Nobili, A., and Signoretti, F.~M. (2023).
\newblock Bank capital requirement shocks: A narrative perspective.
\newblock {\em European Economic Review}, 151:104254.

\bibitem[De~Santis and Van~der Veken, 2026]{de2026deflationary}
De~Santis, R.~A. and Van~der Veken, W. (2026).
\newblock Deflationary financial shocks and inflationary uncertainty shocks: An svar investigation.
\newblock {\em Oxford Bulletin of Economics and Statistics}, 88(1):157--171.

\bibitem[Fanelli and Marsi, 2022]{fanelli2022sovereign}
Fanelli, L. and Marsi, A. (2022).
\newblock Sovereign spreads and unconventional monetary policy in the euro area: A tale of three shocks.
\newblock {\em European Economic Review}, 150:104281.

\bibitem[Fernald, 2014]{fernald2014quarterly}
Fernald, J. (2014).
\newblock A quarterly, utilization-adjusted series on total factor productivity.
\newblock Federal Reserve Bank of San Francisco.

\bibitem[Forni et~al., 2019]{Froni2019}
Forni, M., Gambetti, L., and Sala, L. (2019).
\newblock Strucutral {VAR}s and noninvertible macroeconomic models.
\newblock {\em Journal of Applied Econometrics}, 34(2):221--246.

\bibitem[Furlanetto and Robstad, 2019]{furlanetto2019immigration}
Furlanetto, F. and Robstad, {\O}. (2019).
\newblock Immigration and the macroeconomy: Some new empirical evidence.
\newblock {\em Review of Economic Dynamics}, 34:1--19.

\bibitem[Gambetti et~al., 2023]{gambetti2023agreed}
Gambetti, L., Korobilis, D., Tsoukalas, J., and Zanetti, F. (2023).
\newblock Agreed and disagreed uncertainty.
\newblock {\em arXiv preprint arXiv:2302.01621}.

\bibitem[Giacomini et~al., 2021]{giacomini2021identification}
Giacomini, R., Kitagawa, T., and Read, M. (2021).
\newblock Identification and inference under narrative restrictions.
\newblock {\em arXiv preprint arXiv:2102.06456}.

\bibitem[Giacomini et~al., 2022a]{giacomini2022narrative}
Giacomini, R., Kitagawa, T., and Read, M. (2022a).
\newblock Narrative restrictions and proxies.
\newblock {\em Journal of Business \& Economic Statistics}, 40(4):1415--1425.

\bibitem[Giacomini et~al., 2022b]{giacomini2022narrativerejoinder}
Giacomini, R., Kitagawa, T., and Read, M. (2022b).
\newblock Narrative restrictions and proxies: Rejoinder.
\newblock {\em Journal of Business \& Economic Statistics}, 40(4):1438--1441.

\bibitem[Giannone et~al., 2015]{giannone2015prior}
Giannone, D., Lenza, M., and Primiceri, G.~E. (2015).
\newblock Prior selection for vector autoregressions.
\newblock {\em Review of Economics and Statistics}, 97(2):436--451.

\bibitem[Gilchrist and Zakraj\v{s}ek, 2012]{gilchrist2012credit}
Gilchrist, S. and Zakraj\v{s}ek, E. (2012).
\newblock Credit spreads and business cycle fluctuations.
\newblock {\em American Economic Review}, 102(4):1692--1720.

\bibitem[Hansen and Sargent, 2019]{hansen2019two}
Hansen, L.~P. and Sargent, T.~J. (2019).
\newblock Two difficulties in interpreting vector autoregressions.
\newblock In {\em Rational expectations econometrics}, pages 77--119. CRC Press.

\bibitem[Harrison et~al., 2023]{harrison2023structural}
Harrison, A., Liu, X., and Stewart, S.~L. (2023).
\newblock Structural sources of oil market volatility and correlation dynamics.
\newblock {\em Energy Economics}, 121:106658.

\bibitem[Hauzenberger et~al., 2022]{hauzenberger2022bayesian}
Hauzenberger, N., Huber, F., Koop, G., and Mitchell, J. (2022).
\newblock Bayesian modeling of tvp-vars using regression trees.
\newblock {\em arXiv preprint arXiv:2209.11970}.

\bibitem[Herwartz and Wang, 2023]{herwartz2023point}
Herwartz, H. and Wang, S. (2023).
\newblock Point estimation in sign-restricted svars based on independence criteria with an application to rational bubbles.
\newblock {\em Journal of Economic Dynamics and Control}, 151:104630.

\bibitem[Huber and Feldkircher, 2019]{huber2019adaptive}
Huber, F. and Feldkircher, M. (2019).
\newblock Adaptive shrinkage in bayesian vector autoregressive models.
\newblock {\em Journal of Business \& Economic Statistics}, 37(1):27--39.

\bibitem[Inoue and Kilian, 2022]{inoue2022joint}
Inoue, A. and Kilian, L. (2022).
\newblock Joint bayesian inference about impulse responses in var models.
\newblock {\em Journal of Econometrics}, 231(2):457--476.

\bibitem[Jarocinski and Ma{\'c}kowiak, 2017]{jarocinski2017granger}
Jarocinski, M. and Ma{\'c}kowiak, B. (2017).
\newblock Granger causal priority and choice of variables in vector autoregressions.
\newblock {\em Review of Economics and Statistics}, 99(2):319--329.

\bibitem[Kilian and Zhou, 2020]{kilian2020does}
Kilian, L. and Zhou, X. (2020).
\newblock Does drawing down the us strategic petroleum reserve help stabilize oil prices?
\newblock {\em Journal of Applied Econometrics}, 35(6):673--691.

\bibitem[Kilian and Zhou, 2022]{kilian2022oil}
Kilian, L. and Zhou, X. (2022).
\newblock Oil prices, exchange rates and interest rates.
\newblock {\em Journal of International Money and Finance}, 126:102679.

\bibitem[Korobilis, 2022]{Korobilis2022}
Korobilis, D. (2022).
\newblock A new algorithm for structural restrictions in {B}ayesian vevtor autoregressions.
\newblock {\em European Economic Review}, 148:104241.

\bibitem[Korobilis, 2025]{korobilis2025exploring}
Korobilis, D. (2025).
\newblock Exploring monetary policy shocks with large-scale bayesian vars.
\newblock {\em arXiv preprint arXiv:2505.06649}.

\bibitem[Larsen, 2021]{larsen2021components}
Larsen, V.~H. (2021).
\newblock Components of uncertainty.
\newblock {\em International Economic Review}, 62(2):769--788.

\bibitem[Laumer, 2020]{laumer2020government}
Laumer, S. (2020).
\newblock Government spending and heterogeneous consumption dynamics.
\newblock {\em Journal of Economic Dynamics and Control}, 114:103868.

\bibitem[Lippi and Reichlin, 1993]{lippi1993dynamic}
Lippi, M. and Reichlin, L. (1993).
\newblock The dynamic effects of aggregate demand and supply disturbances: Comment.
\newblock {\em The American Economic Review}, 83(3):644--652.

\bibitem[Lippi and Reichlin, 1994]{lippi1994var}
Lippi, M. and Reichlin, L. (1994).
\newblock Var analysis, nonfundamental representations, blaschke matrices.
\newblock {\em Journal of Econometrics}, 63(1):307--325.

\bibitem[Ludvigson et~al., 2021]{ludvigson2021uncertainty}
Ludvigson, S.~C., Ma, S., and Ng, S. (2021).
\newblock Uncertainty and business cycles: exogenous impulse or endogenous response?
\newblock {\em American Economic Journal: Macroeconomics}, 13(4):369--410.

\bibitem[Maffei-Faccioli and Vella, 2021]{maffei2021does}
Maffei-Faccioli, N. and Vella, E. (2021).
\newblock Does immigration grow the pie? asymmetric evidence from germany.
\newblock {\em European Economic Review}, 138:103846.

\bibitem[Makalic and Schmidt, 2015]{makalic2015simple}
Makalic, E. and Schmidt, D.~F. (2015).
\newblock A simple sampler for the horseshoe estimator.
\newblock {\em IEEE Signal Processing Letters}, 23(1):179--182.

\bibitem[Neri, 2023]{neri2023long}
Neri, S. (2023).
\newblock Long-term inflation expectations and monetary policy in the euro area before the pandemic.
\newblock {\em European Economic Review}, 154:104426.

\bibitem[Pfarrhofer and Stelzer, 2024]{pfarrhofer2024high}
Pfarrhofer, M. and Stelzer, A. (2024).
\newblock High-frequency and heteroskedasticity identification in multicountry models: Revisiting spillovers of monetary shocks.
\newblock Technical report, arXiv. org.

\bibitem[Plagborg-M{\o}ller, 2022]{plagborg2022discussion}
Plagborg-M{\o}ller, M. (2022).
\newblock Discussion of “narrative restrictions and proxies” by raffaella giacomini, toru kitagawa, and matthew read.
\newblock {\em Journal of Business \& Economic Statistics}, 40(4):1434--1437.

\bibitem[Pr{\"u}ser, 2024]{pruser2024large}
Pr{\"u}ser, J. (2024).
\newblock A large non-gaussian structural var with application to monetary policy.
\newblock {\em arXiv preprint arXiv:2412.17598}.

\bibitem[Redl, 2020]{redl2020uncertainty}
Redl, C. (2020).
\newblock Uncertainty matters: Evidence from close elections.
\newblock {\em Journal of International Economics}, 124:103296.

\bibitem[Reichlin et~al., 2023]{reichlin2023monetary}
Reichlin, L., Ricco, G., and Tarb{\'e}, M. (2023).
\newblock Monetary--fiscal crosswinds in the european monetary union.
\newblock {\em European Economic Review}, 151:104328.

\bibitem[R{\"u}th and Van~der Veken, 2023]{ruth2023monetary}
R{\"u}th, S.~K. and Van~der Veken, W. (2023).
\newblock Monetary policy and exchange rate anomalies in set-identified svars: Revisited.
\newblock {\em Journal of Applied Econometrics}, 38(7):1085--1092.

\bibitem[Uhlig, 2005]{uhlig2005effects}
Uhlig, H. (2005).
\newblock What are the effects of monetary policy on output? {R}esults from an agnostic identification procedure.
\newblock {\em Journal of Monetary Economics}, 52(2):381--419.

\bibitem[Zeev, 2018]{zeev2018can}
Zeev, N.~B. (2018).
\newblock What can we learn about news shocks from the late 1990s and early 2000s boom-bust period?
\newblock {\em Journal of Economic Dynamics and Control}, 87:94--105.

\bibitem[Zhou, 2020]{zhou2020refining}
Zhou, X. (2020).
\newblock Refining the workhorse oil market model.
\newblock {\em Journal of Applied Econometrics}, 35(1):130--140.

\end{thebibliography}

\emptythanks 

\newcounter{footnotecounter}
    \setcounter{footnotecounter}{\value{footnote}}
    \setcounter{footnote}{0}



\clearpage
    \title{ONLINE APPENDIX - Sharpening Identification in Large Structural VARs Using Narrative Restrictions}
\begin{titlepage}
\setlength{\droptitle}{-1cm} 
\maketitle
\begin{abstract}
	\begin{singlespace}
 \noindent
 This is the online appendix for ``Sharpening Identification in Large Structural VARs Using Narrative Restrictions''. The reader is referred to the paper for more detailed information.
	\end{singlespace}
\end{abstract}
\bigskip

\noindent \textbf{Keywords:} Structural Identification, Sign Restrictions, Narrative Restrictions, Large BVARs  \bigskip
 
\noindent \textbf{JEL classification:} C11, C32, C55, E50	 \bigskip
 
\end{titlepage}
\pagenumbering{arabic}

\appendix

\renewcommand{\thesection}{Online Appendix \Alph{section}}
\setcounter{table}{0}
\renewcommand{\thetable}{\Alph{section}.\arabic{table}}
\setcounter{figure}{0}
\renewcommand{\thefigure}{\Alph{section}.\arabic{figure}}

\newpage

\onehalfspacing

\section{Remaining Prior Distributions}

We complete our model specification by describing the further prior distributions.
For the variance terms of the noise we assume $\sigma_{j}^2 \sim \mathcal{IG}(\alpha_0,\beta_0)$.  In our empirical application, we set $\boldsymbol{l}_{0,i}= \boldsymbol{0}$, $ \boldsymbol{V}_{\boldsymbol{l}_i}=10  \times \boldsymbol{I}_r$ and $\alpha_0=\beta_0=0$. In high-dimensional settings such as large VARs, it is important to impose shrinkage prior to mitigate the risk of overfitting. We follow \cite{Korobilis2022} and use the horseshoe prior proposed by \cite{carvalho2010horseshoe}. Let $\boldsymbol{\beta}_i$ the VAR coefficients in the $i-$th equation, $i=1,\dots,n$ and  $\beta_{i,j}$, the $j-$th coefficient in the $i-$th equation. Consider the prior for $\beta_{i,j}$, $i=1,\dots,n$ and $j=2,\dots,k$:
\begin{align}
\beta_{i,j}|\lambda_i, \psi_{i,j} &\sim N(0,\lambda_{i}\psi_{i,j}),\\
\sqrt{\psi_{i,j}}&\sim  C^+(0,1),\\
 \sqrt{\lambda_i}&\sim C^+(0,1),
\end{align}
where $C^+(0,1)$ denotes the standard half-Cauchy distribution. The  hyperparameter $\lambda_i$ is the global variance component that are common to all elements in  $\boldsymbol{\beta}_i$ , whereas each $\psi_{i,j}$ is a local variance component specific to the coefficients $\beta_{i,j}$. To facilitate sampling, we follow \cite{makalic2015simple} and use the following latent variables representations of the half-Cauchy distributions:
\begin{align}
(\psi_{i,j}|z_{\psi_{i,j}})&\sim  \mathcal{IG}(1/2,1/z_{\psi_{i,j}}),\quad z_{\psi_{i,j}}\sim  \mathcal{IG}(1/2,1),\\
(\lambda_i|z_{\lambda_i})&\sim  \mathcal{IG}(1/2,1/z_{\lambda_i}),\quad z_{\lambda_i}\sim \mathcal{IG}(1/2,1),
\end{align}
for $i=1,\dots,n$ and $j=2,\dots,k$.

\section{Gibbs Sampler}

In this section, we develop an efficient posterior sampler to estimate the model. To impose the inequality restrictions on the factors or factor loadings we use the sampler from \cite{botev2017normal}. In contrast to existing rejecting sampling approaches (see, e.g. \cite{antolin2018narrative} or \cite{giacomini2021identification}) each proposed draw is accepted with certainty. Posterior samples are obtained sequentially from the conditional distributions:
\begin{enumerate}
\item $p(\boldsymbol{f}|\boldsymbol{y},\boldsymbol{\beta}, \boldsymbol{L},\boldsymbol{\Sigma}, \boldsymbol{\lambda}, \boldsymbol{\psi}, \boldsymbol{z}_{\lambda},\boldsymbol{z}_{\boldsymbol{\psi}})=p(\boldsymbol{f}|\boldsymbol{y}, \boldsymbol{\beta}, \boldsymbol{L}, \boldsymbol{W}, \boldsymbol{\Sigma})$;
\item $p( \boldsymbol{L}|\boldsymbol{y}\boldsymbol{\beta},\boldsymbol{f},\boldsymbol{\Sigma}, \boldsymbol{\lambda}, \boldsymbol{\psi}, \boldsymbol{z}_{\lambda},\boldsymbol{z}_{\boldsymbol{\psi}})=\prod_{i=1}^np(\boldsymbol{l}_i|\boldsymbol{y}_i, \boldsymbol{\beta}_i,\boldsymbol{f}, \boldsymbol{\sigma}^2_i)$
\item $p(\boldsymbol{\beta}|\boldsymbol{y}, \boldsymbol{L}, \boldsymbol{f},\boldsymbol{\Sigma}, \boldsymbol{\lambda}, \boldsymbol{\psi}, \boldsymbol{z}_{\lambda},\boldsymbol{z}_{\boldsymbol{\psi}})=\prod_{i=1}^n p(\boldsymbol{\beta}_i|\boldsymbol{y}_i,\boldsymbol{l}_i,\boldsymbol{f}, \boldsymbol{\sigma}^2_i)$
\item $p(\boldsymbol{\Sigma}|\boldsymbol{y},\boldsymbol{\beta}, \boldsymbol{L},\boldsymbol{f}, \boldsymbol{\lambda}, \boldsymbol{\psi}, \boldsymbol{z}_{\lambda},\boldsymbol{z}_{\boldsymbol{\psi}})=\prod_i^n p(\sigma_i^2|\boldsymbol{y}_i,\boldsymbol{f}_i,\boldsymbol{l}_i,\boldsymbol{\beta}_i) $;
\item $p(\boldsymbol{\lambda}|\boldsymbol{y},\boldsymbol{\beta}, \boldsymbol{L},\boldsymbol{f},\boldsymbol{v}, \boldsymbol{\psi}, \boldsymbol{z}_{\lambda},\boldsymbol{z}_{\boldsymbol{\psi}})=\prod_{l=1}^n p(\lambda_i|\boldsymbol{\beta}, \boldsymbol{\psi}, z_{\lambda_i})$;
\item $p(\boldsymbol{\psi}|\boldsymbol{y},\boldsymbol{\beta}, \boldsymbol{L},\boldsymbol{f},\boldsymbol{\Sigma}, \boldsymbol{\lambda}, \boldsymbol{z}_{\lambda},\boldsymbol{z}_{\boldsymbol{\psi}})=\prod_{i=1}^n\prod_{j=2}^k p(\psi_{i,j}|\beta_{i,j}, \lambda_i,z_{\psi_{i,j}})$;
\item $p( \boldsymbol{z}_{\lambda}|\boldsymbol{y},\boldsymbol{\beta}, \boldsymbol{L},\boldsymbol{f},\boldsymbol{\Sigma}, \boldsymbol{\lambda}, \boldsymbol{\psi},\boldsymbol{z}_{\boldsymbol{\psi}})=\prod_{i=1}^n p(z_{\lambda_i}|\lambda_i)$;
\item $p(\boldsymbol{z}_{\boldsymbol{\psi}}|\boldsymbol{y},\boldsymbol{\beta}, \boldsymbol{L},\boldsymbol{f},\boldsymbol{\Sigma}, \boldsymbol{\lambda}, \boldsymbol{\psi}, \boldsymbol{z}_{\lambda})=\prod_{i=1}^n\prod_{j=2}^k p(z_{ \psi_{i,j}}| \psi_{i,j})$,
\end{enumerate}
with $\boldsymbol{y}_i=(y_{i,1},\dots,y_{i,T})'$ be a $T\times 1$ vector of observations of the $i-$th variable.

\textbf{Step 1} First, we sample $\boldsymbol{f}_t$ for $t=1,\dots,T$. We can use standard regression results (see, e.g., \cite{chan2019bayesian}) to obtain 
\begin{equation}
(\boldsymbol{f}_t|\boldsymbol{y},\boldsymbol{\beta}, \boldsymbol{L})\sim N(\hat{\boldsymbol{f}}_t,\boldsymbol{K}_{\boldsymbol{f}}^{-1}),
\end{equation}
where 
\begin{equation}
\boldsymbol{K}_{\boldsymbol{f}}^{-1}=(\boldsymbol{I}_r+\boldsymbol{L}' \boldsymbol{\Sigma} \boldsymbol{L})^{-1}, \quad \hat{\boldsymbol{f}}_t=\boldsymbol{K}_{\boldsymbol{f}}(\boldsymbol{L} \boldsymbol{\Sigma}^{-1}(\boldsymbol{y}_t- (\boldsymbol{I}_n  \otimes \boldsymbol{x}'_t)\boldsymbol{\beta} )).
\end{equation}
We use the efficient truncated Normal generator provided by \cite{botev2017normal} in order to sample the factors.

\textbf{Step 2} Second, we sample $\boldsymbol{L}$. Given the latent factors $\boldsymbol{f}$, the VAR becomes $n$ unrelated regressions and we can sample $\boldsymbol{L}$ equation by equation. Remember that $\boldsymbol{\beta}_i$ and $\boldsymbol{l}_i$ denote, respectively, the VAR coefficients and the factor loadings in the $i-$th equation. Then, the $i-$th equation of the VAR can be expressed as
\begin{equation}
\boldsymbol{y}_i=\boldsymbol{X}_i \boldsymbol{\beta}_i+\boldsymbol{F} \boldsymbol{l}_i +\boldsymbol{v}
\label{eqbyeq}
\end{equation}
where $\boldsymbol{F}=(\boldsymbol{f}_1,\dots,\boldsymbol{f}_r)$ the $ T\times r$ matrix of factors with $ \boldsymbol{f}_i=(f_{i,1},\dots,f_{i,T})'$. The vector of noise $\boldsymbol{v}=(v_{i,1},\dots,v_{i,T})'$ is distributed as $N(\boldsymbol{0},\boldsymbol{I}_T \sigma_i^2)$. 

Then using standard linear regression results, we get
\begin{equation}
(\boldsymbol{l}_i|\boldsymbol{y}_i,\boldsymbol{f}, \boldsymbol{\beta}_i, \sigma_i^2)\sim N( \hat{\boldsymbol{l}_i},\boldsymbol{K}^{-1}_{\boldsymbol{l}_i})\boldsymbol{1} (\ell_{t}^{f} < \boldsymbol{R}_{t}^{f} \boldsymbol{f}_{t}< \upsilon_{t}^{f})\boldsymbol{1} (\ell_{it}^{Lf} < \boldsymbol{R}_{it}^{Lf} (\boldsymbol{l}_i' \odot \boldsymbol{f}_{t})< \upsilon_{it}^{Lf},
\end{equation}
where
\begin{align*}
\boldsymbol{K}_{\boldsymbol{l}_i}=\boldsymbol{V}^{-1}_{\boldsymbol{l}_i}+\boldsymbol{\sigma}^{-2}_i \boldsymbol{F}'\boldsymbol{F}, \quad  \hat{\boldsymbol{l}}_i=\boldsymbol{K}^{-1}_{\boldsymbol{l}_i}(\boldsymbol{V}^{-1}_{\boldsymbol{l}_i}\boldsymbol{l}_{0,i}+\sigma^{-2}_i \boldsymbol{F} (\boldsymbol{y}_i-\boldsymbol{X}_i \boldsymbol{\beta}_i)).
\end{align*}
We use the efficient truncated Normal generator provided by \cite{botev2017normal} in order to sample $\boldsymbol{l}_i$.

\textbf{Step 3} Third, we sample $\boldsymbol{\beta}$ equation by equation based on (\ref{eqbyeq}). Equation by equation estimation simplifies the estimation and allows for the estimation with a large number of variables. Again, 
using standard linear regression results, we get
\begin{equation}
(\boldsymbol{\beta}_i|\boldsymbol{y}_i,\boldsymbol{f}, \boldsymbol{l}_i, \sigma_i^2)\sim N( \hat{\boldsymbol{l}_i},\boldsymbol{K}^{-1}_{\boldsymbol{\beta}_i}),
\end{equation}
where
\begin{align*}
\boldsymbol{K}_{\boldsymbol{\beta}_i}=\boldsymbol{V}^{-1}_{\boldsymbol{\beta}_i}+\boldsymbol{\sigma}^{-2}_i \boldsymbol{X}_i'\boldsymbol{X}_i, \quad  \hat{\boldsymbol{\beta}}_i=\boldsymbol{K}^{-1}_{\boldsymbol{\beta}_i}(\boldsymbol{V}^{-1}_{\boldsymbol{l}_i}\boldsymbol{\beta}_{0,i}+\sigma^{-2}_i \boldsymbol{X}_i (\boldsymbol{y}_i-\boldsymbol{F}\boldsymbol{l}_i)).
\end{align*}

\textbf{Step 4} Next, we sample $\sigma^2_{i}$ for $i=1,\dots,n$. Given $\boldsymbol{f}$ the model reduces to $n$ independent linear regressions. Therefore, we can use standard regression results (see, e.g., \cite{chan2019bayesian}) to obtain
\begin{equation}
(\sigma_{i}^2|\boldsymbol{y}, \boldsymbol{f}, \boldsymbol{L})\sim \mathcal{IG}\left(\alpha_0+\frac{T}{2},\beta_0+0.5\sum_{t=1}^T(y_{it}-\boldsymbol{X}_{it}\boldsymbol{\beta}_i -\boldsymbol{l}_i\boldsymbol{f}_t)^2\right).  
\end{equation}

\textbf{Step 5} Lastly, we sample the hyperparameter $\lambda_{i}$ and $\psi_{i,j}$ from our shrinkage prior for the VAR coefficients as well as the mixing variables $z_{\lambda_{i}}$ and $z_{\psi_{i,j}}$. Using the latent variable representation of the half Cauchy distribution, we obtain
\begin{align*}
p(\psi_{i,j}|\beta_{i,j}, \lambda_{i}, z_{\psi_{i,j}})&\propto \psi_{i,j}^{\frac{1}{2}} \text{e}^{-\frac{1}{2\lambda_{i}\psi_{i,j}}(\beta_{i,j})^2  }\times \psi^{-\frac{3}{2}}\text{e}^{-\frac{1}{\psi_{i,j}z_{\psi_{i,j}}}}\\
&=\psi_{i,j}^{-2}\text{e}^{-\frac{1}{\psi_{i,j}}\left(\frac{1}{z_{\psi_{i,j}}}+\frac{(\beta_{i,j})^2}{2\lambda_{i}}\right)},
\end{align*}
which is the kernel of the following inverse-gamma distribution:
\begin{equation}
(\psi_{i,j}|\beta_{i,j}, \lambda_{i}, z_{\psi_{i,j}})\sim \mathcal{IG}\left(1, \frac{1}{z_{\psi_{i,j}}}+\frac{(\beta_{i,j})^2}{2\lambda_{i}}\right).
\end{equation}
Furthermore,
\begin{align*}
p(\lambda_i|\boldsymbol{\beta}, \boldsymbol{\psi}, z_{\lambda_i})&\propto \prod_{i} \lambda_i^{-\frac{1}{2}} \text{e}^{-\frac{1}{2\lambda_i \psi_{i,j}}(\beta_{i,j})^2   }\times \lambda_i^{-\frac{3}{2}} \text{e}^{-\frac{1}{\lambda_i z_{\lambda_i}}  },\\
&=\lambda_i^{-\left(\frac{k}{2}+1 \right)}
\text{e}^{-\frac{1}{\lambda_i}\left(\frac{1}{z_{\lambda_i}} +\sum_{j=2}^{k}\frac{(\beta_{i,j})^2}{2\psi_{i,j}}  \right)   },
\end{align*}
which is the kernel of the following inverse-gamma distribution:
\begin{equation}
(\lambda_i|\boldsymbol{\beta}, \boldsymbol{\psi}, z_{\psi_{\lambda_i}})\sim \mathcal{IG}\left(\frac{k}{2}, \frac{1}{z_{\lambda_i} }+\sum_{j=2}^{k}\frac{(\beta_{i,j})^2}{2\psi_{i,j}}\right).
\end{equation}
Finally, we sample the latent variables $\boldsymbol{z}_{\boldsymbol{\psi}}$ and $\boldsymbol{z}_{\boldsymbol{\lambda}}$. In particular, $z_{\psi_{i,j}}\sim \mathcal{IG}(1,1+\psi_{i,j}^{-1})$ for $i=1,\dots,n$ and $j=2,\dots,k$. Similarly, we have $z_{\lambda_{l}}\sim  \mathcal{IG}(1,1+\lambda_i^{-1})$ for $i=1,\dots,n$.

\clearpage
\section{Data - Empirical Application} \label{data_set}

Table \ref{tab_data} includes all variables used in the empirical application analysis, together with a description of the variables, the transformation code and the data source.

The transformation codes used in column "T" are 1 - no transformation and 2 - log levels.The column denoted with "S" refers to the data source and uses the following source codes: 1 - Federal Reserve Bank of St. Louis (FRED), 2 - \cite{fernald2014quarterly}, 3 - Refinitiv, 4 -  \cite{gilchrist2012credit} and 5 - \cite{baker2016measuring}.

\begin{table}[H]
\centering
\caption{Variable description, transformation and data sources}
\scriptsize
\setlength{\tabcolsep}{4pt}      
\renewcommand{\arraystretch}{0.85} 
\rowcolors{2}{gray!10}{white}
\begin{tabular}{l >{\raggedright\arraybackslash\linespread{0.95}\selectfont}p{12cm} c c}
\toprule
\textbf{Variable} & \textbf{Description} & \textbf{T} & \textbf{S} \\
\midrule
GDP & Real Gross Domestic Product, Seasonally Adjusted Annual Rate & 2 & 1 \\
PCE & Real Personal Consumption Expenditures, Seasonally Adjusted Annual Rate, Sum & 2 & 1 \\
Residential investment & Private Residential Fixed Investment, Seasonally Adjusted Annual Rate & 2 & 1 \\
Nonresidential investment & Private Nonresidential Fixed Investment, Seasonally Adjusted Annual Rate, Sum & 2 & 1 \\
Exports & Real Exports of Goods and Services, Seasonally Adjusted Annual Rate & 2 & 1 \\
Imports & Real Imports of Goods and Services, Seasonally Adjusted Annual Rate & 2 & 1 \\
Government spending & Real Government Consumption Expenditures and Gross Investment, Seasonally Adjusted Annual Rate  & 2 & 1 \\
Fed. buget surplus/deficit & (Federal government total receipts, Seasonally Adjusted Annual Rate minus Federal government total expenditures, Seasonally Adjusted Annual Rate) divided by Gross Domestic Product, Seasonally Adjusted Annual Rate & 1 & 1 \\
Fed. tax receipts & Federal Government Current Receipts, Seasonally Adjusted Annual Rate & 1 & 1 \\
GDP deflator & Gross Domestic Product: Implicit Price Deflator, Index 2017=100, Seasonally Adjusted  & 2 & 1 \\
PCE index & Personal Consumption Expenditures: Chain-type Price Index, Index 2017=100, Seasonally Adjusted & 2 & 1 \\
PCE index less F\&E & Personal Consumption Expenditures Excluding Food and Energy (Chain-Type Price Index), Index 2017=100, Seasonally Adjusted & 2 & 1 \\
CPI index & Consumer Price Index for All Urban Consumers: All Items in U.S. City Average, Index 1982-1984=100, Seasonally Adjusted & 2 & 1 \\
CPI index less F\&E & Consumer Price Index for All Urban Consumers: All Items Less Food and Energy in U.S. City, Index 1982-1984=100, Seasonally Adjusted Average & 2 & 1 \\
Hourly wage & Nonfarm Business Sector: Real Hourly Compensation for All Workers, Index 2017=100, Seasonally Adjusted & 2 & 1 \\
Labor productivity & Nonfarm Business Sector: Labor Productivity (Output per Hour) for All Workers Index 2017=100, Seasonally Adjusted & 2 & 1 \\
Utilization-adjusted TFP & Utilization-adjusted Total Factor Productivity, Cumulative Sum based on \cite{fernald2014quarterly} & 2 & 2 \\
Employment & All Employees, Total Nonfarm, Seasonally Adjusted & 2 & 1 \\
Unemployment rate & Unemployment Rate, Seasonally Adjusted, Average & 1 & 1 \\
Industrial production index & Industrial Production: Total Index, Index 2017=100, Seasonally Adjusted & 2 & 1 \\
Capacity utilization & Capacity Utilization: Total Index, Seasonally Adjusted & 1 & 1 \\
Housing starts & New Privately-Owned Housing Units Started: Total Units, Seasonally Adjusted Annual Rate, Average & 2 & 1 \\
Disposable income & Real Disposable Personal Income, Seasonally Adjusted Annual Rate, Sum & 2 & 1 \\
Consumer sentiment & University of Michigan: Consumer Sentiment, Not Seasonally Adjusted, Average & 2 & 1 \\
Fed funds rate & Federal Funds Effective Rate, Not Seasonally Adjusted, Average & 1 & 1 \\
3-month T-note rate & 3-Month Treasury Bill Secondary Market Rate, Discount Basis, Not Seasonally Adjusted, Average & 1 & 1 \\
2-year T-note rate & Market Yield on U.S. Treasury Securities at 2-Year Constant Maturity, Quoted on an Investment Basis, Not Seasonally Adjusted, Average & 1 & 1 \\
5-year T-note rate & Market Yield on U.S. Treasury Securities at 5-Year Constant Maturity, Quoted on an Investment Basis, Not Seasonally Adjusted, Average  & 1 & 1 \\
10-year T-note rate & Market Yield on U.S. Treasury Securities at 10-Year Constant Maturity, Quoted on an Investment Basis, Not Seasonally Adjusted, Average & 1 & 1 \\
Prime rate & Bank Prime Loan Rate, Not Seasonally Adjusted, Average & 1 & 1 \\
Aaa corporate bond yield & Moody's Seasoned Aaa Corporate Bond Yield, Not Seasonally Adjusted, Average & 1 & 1 \\
Baa corporate bond yield & Moody's Seasoned Baa Corporate Bond Yield, Not Seasonally Adjusted, Average & 1 & 1 \\
Trade-weighted US index & Trade Weighted U.S. Dollar Index: Major Currencies, Goods, Index Mar 1973=100,
Not Seasonally Adjusted, Average \& Nominal Advanced Foreign Economies U.S. Dollar Index, , Index Mar 2006=100,
Not Seasonally Adjusted, Average & 2 & 1 \\
S\&P500 & S\&P 500 Composite - Price Index, Closing, Not Seasonally Adjusted, Average & 2 & 3 \\
Spot oil price & Spot Crude Oil Price: West Texas Intermediate (WTI), Not Seasonally Adjusted, Average & 2 & 1 \\
Liquidity & Nonfinancial Corporate Business; Liquid Assets (Broad Measure), Not Seasonally Adjusted divided by Business Sector: Value-Added Output Price Deflator for All Workers, Index 2017=100, Seasonally Adjusted & 2 & 1 \\
GZ spread & Credit spread based on \cite{gilchrist2012credit} & 1 & 4 \\
EBP & Excess bond premium based \cite{gilchrist2012credit} & 1 & 4 \\
EPU & Economic Policy Uncertainty Index based \cite{baker2016measuring} & 2 & 5 \\
\bottomrule
\end{tabular}
\label{tab_data}
\vspace{0.3em}
\begin{minipage}{0.9\textwidth}
\footnotesize
\textit{Notes:} This table lists the 39 variables used in the empirical application, along with their descriptions and transformation and source codes.
\end{minipage}
\end{table}

\clearpage
\section{Additional Empirical Results} \label{app_add_emp_results}

\begin{table}[H]
\centering
\renewcommand{\arraystretch}{0.9} 
\begin{tabular}{lcccccccccc}
\hline
\textbf{Quarter} & \textbf{Dem} & \textbf{Inv} & \textbf{Fin} & \textbf{Mon} & \textbf{Gov} & \textbf{Tec} & \textbf{Lab} & \textbf{Wag} & \textbf{Oil} & \textbf{Con}  \\
\hline
1978Q4 &  &  &  &  &  &  &  &  & $<0$ &   \\
1980Q1 &  &  &  &  & $>0$ &  &  &  &  &    \\
1980Q3 &  &  &  &  &  &  &  &  & $<0$ &    \\
1980Q4 &  &  &  &  &  &  &  &  & $<0$ &    \\
1990Q3 &  &  &  &  &  &  &  &  & $<0$ &    \\
2001Q3 &  &  &  &  & $>0$ &  &  &  &  &   \\
2002Q4 &  &  &  &  &  &  &  &  & $<0$ &    \\
2003Q1 &  &  &  &  &  &  &  &  & $<0$ &    \\
2006Q1 & $>0$ &  &  &  & &  &  &  &  &    \\
2007Q3 &  &  & $>0$ &  &  &  &  &  &  &  \\
2008Q3 &  &  & $>0$ &  &  &  &  &  &  &   \\
2008Q4 &  &  & $>0$ &  &  &  &  &  &  &   \\
2011Q1 &  &  &  &  &  &  &  &  & $<0$ &    \\
2011Q3 &  &  &  &  &  &  &  &  &  &   \\
\hline
\end{tabular}

\vspace{0.5em}
\parbox{\textwidth}{\scriptsize
\textit{Notes:} The table reports shock sign restrictions imposed in the extended model.
Entries $>0$ ($<0$) indicate that the corresponding shock is assumed to be positive (negative)
in the respective quarter. The abbreviations correspond as follows: Dem: demand; Inv: investment; Fin: financial; Mon: monetary policy; Gov: government spending;
Tec: technology; Lab: labor supply; Wag: wage bargaining; Oil: oil supply; Con: consumer sentiment.}
\caption{Narrative shock sign restrictions in the \nth{2} extended model.}
\label{tab:shock-quarter-restrictions_alternative}
\end{table}

\begin{figure}[H]
\centering
\setlength{\tabcolsep}{2pt}
\renewcommand{\arraystretch}{1}

\begin{tabular}{ccc}
GDP & Investment & Imports \\

\includegraphics[
    width=0.32\textwidth,
    trim=1mm 1mm 1mm 1mm,
    clip
]{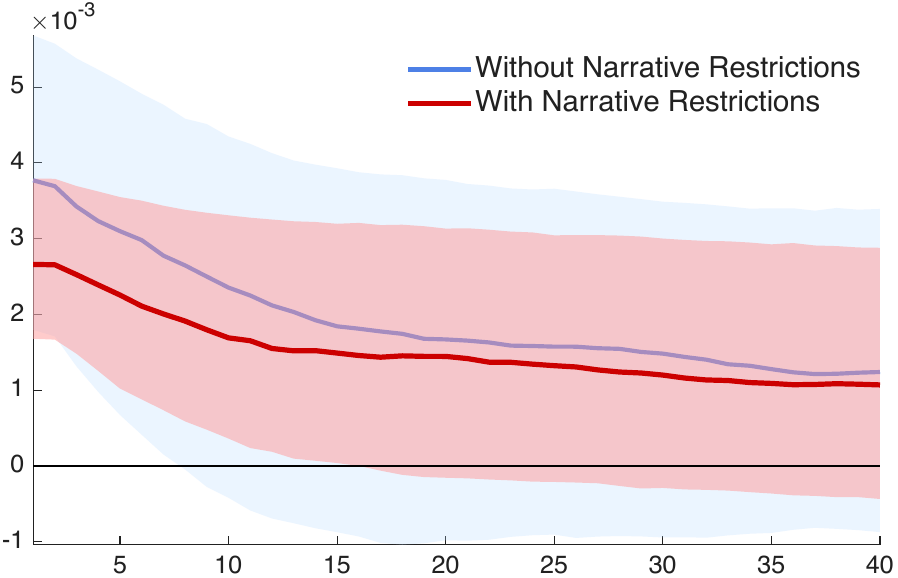}
&
\includegraphics[
    width=0.32\textwidth,
    trim=1mm 1mm 1mm 1mm,
    clip
]{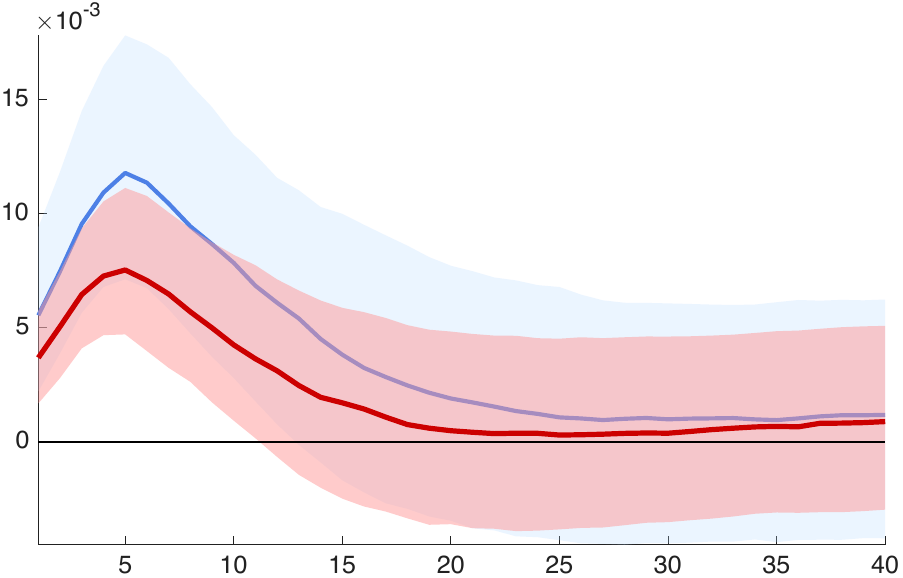}
&
\includegraphics[
    width=0.32\textwidth,
    trim=1mm 1mm 1mm 1mm,
    clip
]{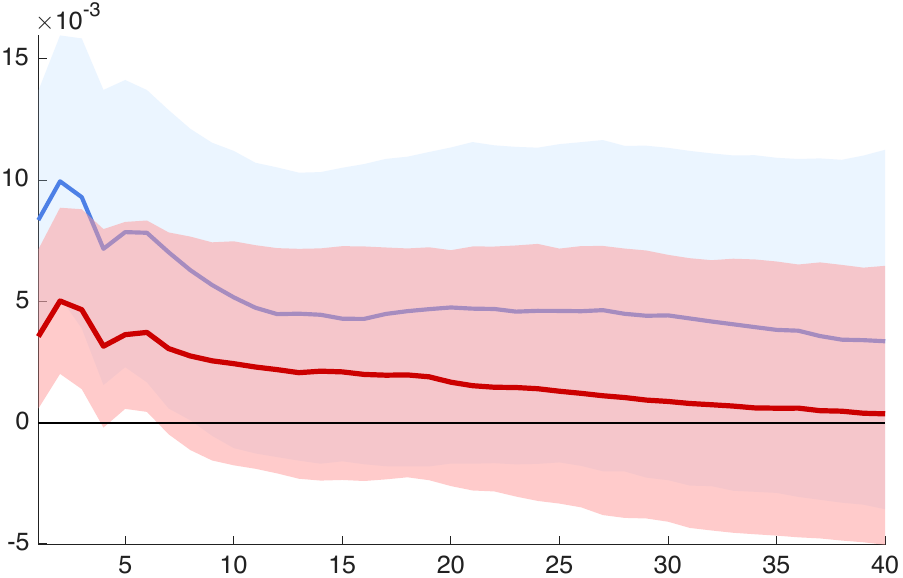}

\\[0.4cm]

GDP Deflator & Unemployment & Disposable Income \\

\includegraphics[
    width=0.32\textwidth,
    trim=1mm 1mm 1mm 1mm,
    clip
]{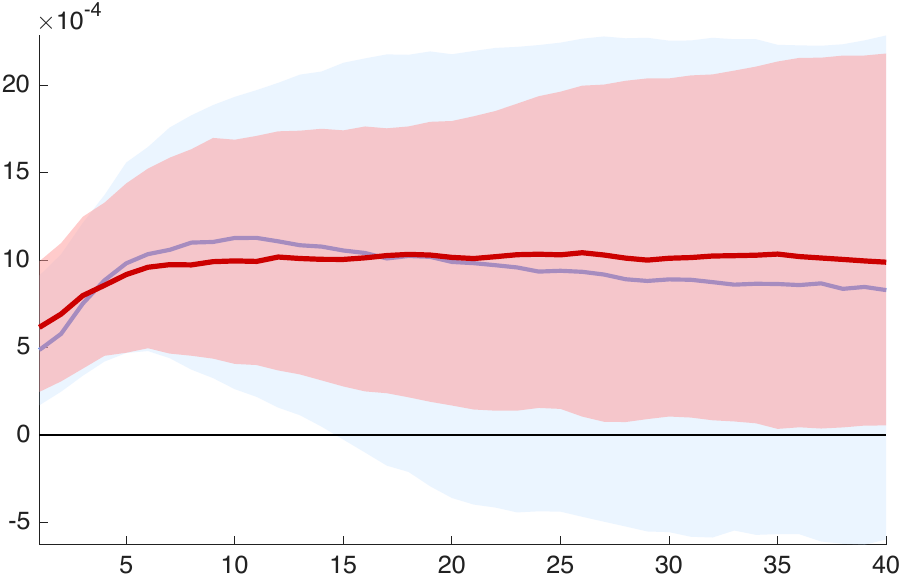}
&
\includegraphics[
    width=0.32\textwidth,
    trim=1mm 1mm 1mm 1mm,
    clip
]{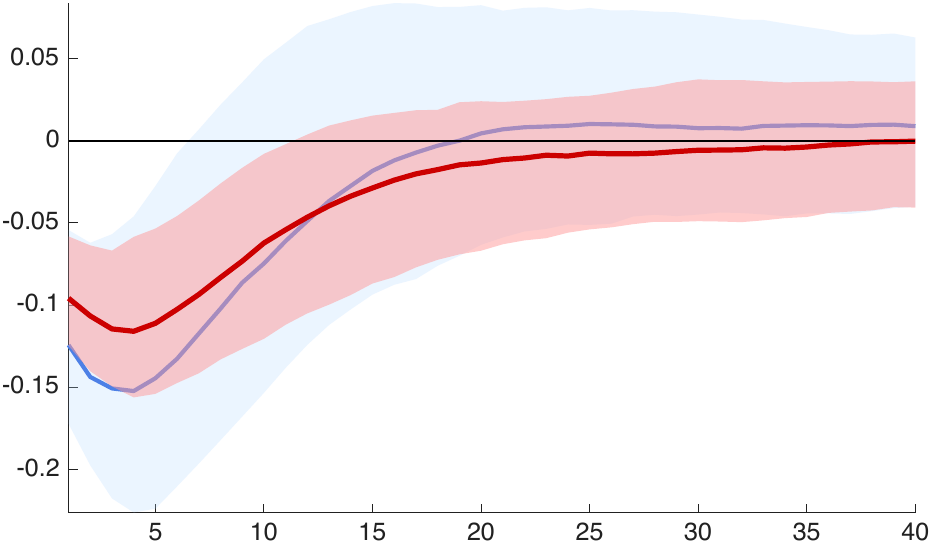}
&
\includegraphics[
    width=0.32\textwidth,
    trim=1mm 1mm 1mm 1mm,
    clip
]{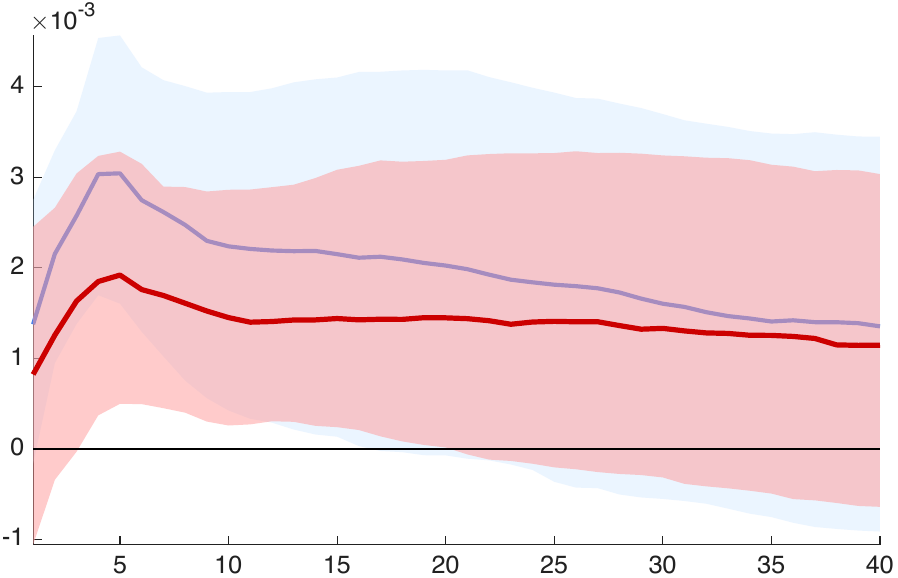}

\end{tabular}
\vspace{-0.25cm} 
\caption{Selected IRFs to Favorable Demand Shock.} 
\label{fig2_demand_shock}
\vspace{0.05cm} 
\begin{minipage}{0.98\textwidth}
\footnotesize
\textit{Notes:} Impulse responses of selected economic indicators.
Blue: 68\% credible sets (no narrative restrictions).
Red: median and 68\% bands (with narrative restrictions).
\end{minipage}
\end{figure}

\begin{figure}[H]
\centering
\setlength{\tabcolsep}{2pt}
\renewcommand{\arraystretch}{1}

\begin{tabular}{ccc}
GDP & Investment & Imports \\

\includegraphics[
    width=0.32\textwidth,
    trim=1mm 1mm 1mm 1mm,
    clip
]{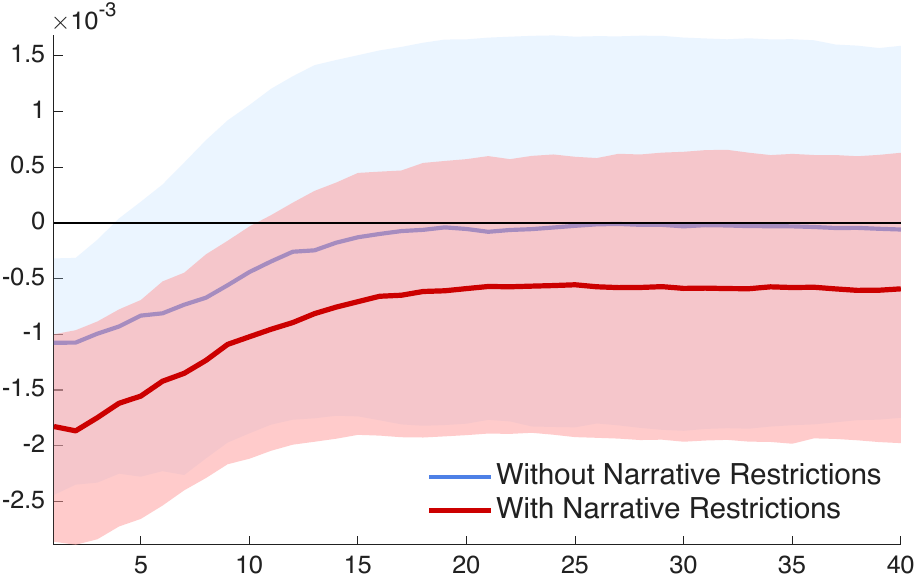}
&
\includegraphics[
    width=0.32\textwidth,
    trim=1mm 1mm 1mm 1mm,
    clip
]{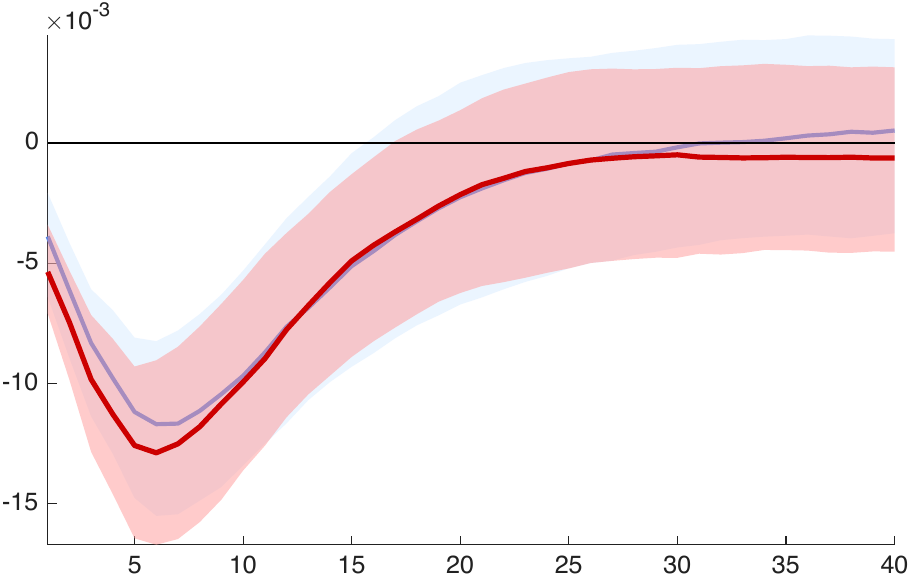}
&
\includegraphics[
    width=0.32\textwidth,
    trim=1mm 1mm 1mm 1mm,
    clip
]{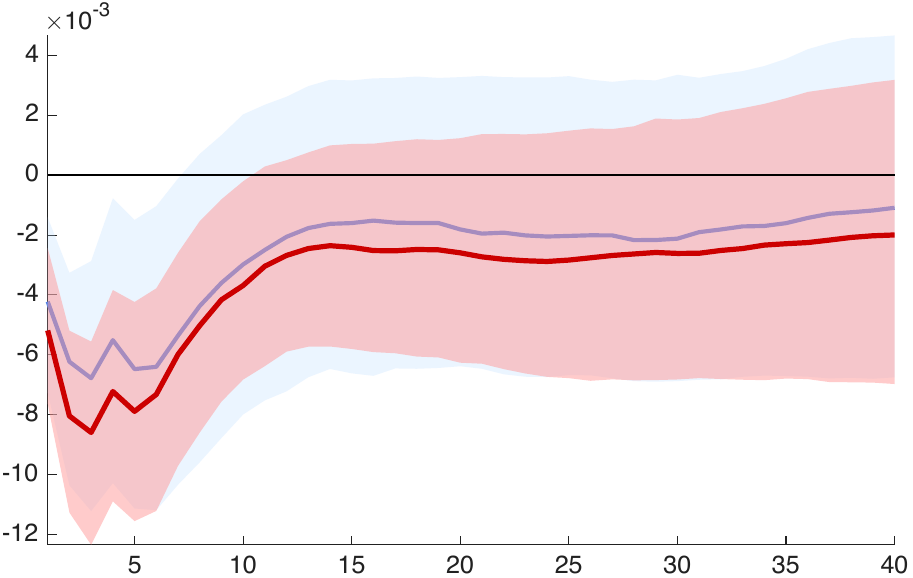}

\\[0.4cm]

GDP Deflator & Unemployment & IP \\

\includegraphics[
    width=0.32\textwidth,
    trim=1mm 1mm 1mm 1mm,
    clip
]{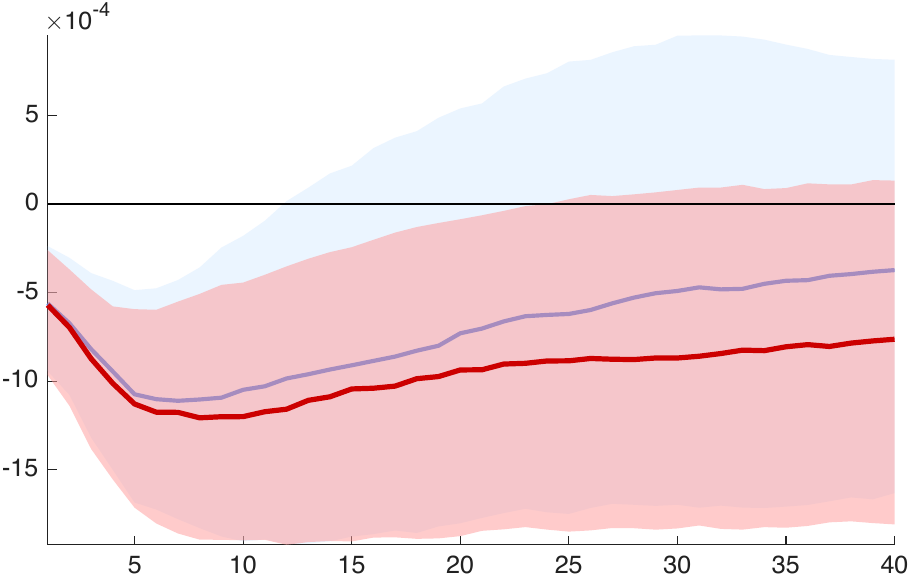}
&
\includegraphics[
    width=0.32\textwidth,
    trim=1mm 1mm 1mm 1mm,
    clip
]{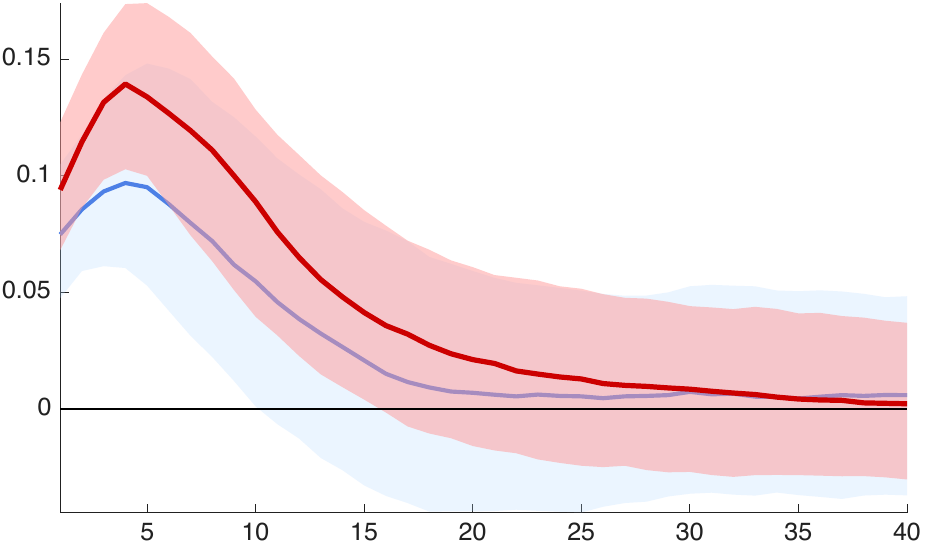}
&
\includegraphics[
    width=0.32\textwidth,
    trim=1mm 1mm 1mm 1mm,
    clip
]{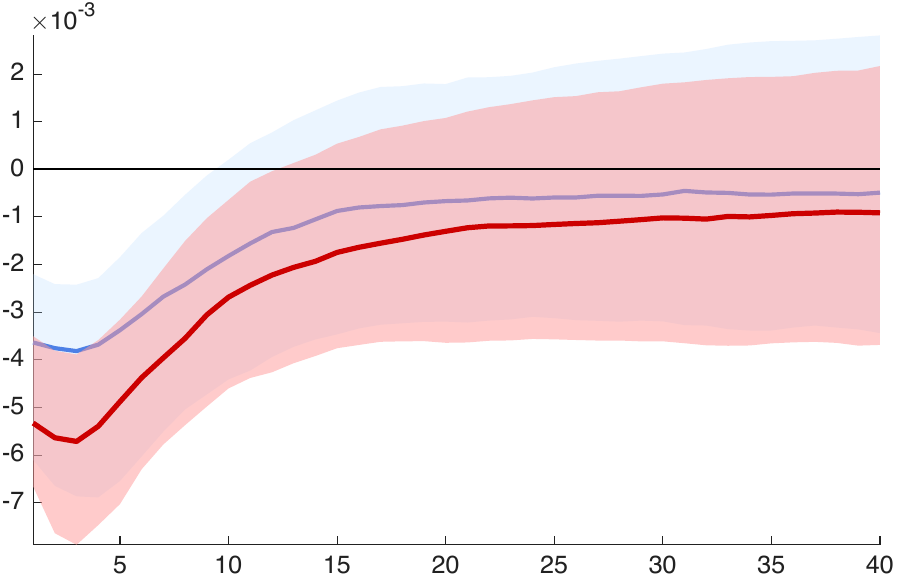}

\end{tabular}
\vspace{-0.25cm} 
\caption{Selected IRFs to Adverse Financial Shock.} 
\label{fig2_financial_shock}
\vspace{0.05cm} 
\begin{minipage}{0.98\textwidth}
\footnotesize
\textit{Notes:} Impulse responses of selected economic indicators.
Blue: 68\% credible sets (no narrative restrictions).
Red: median and 68\% bands (with narrative restrictions).
\end{minipage}
\end{figure}

\begin{figure}[H]
\centering
\setlength{\tabcolsep}{2pt}
\renewcommand{\arraystretch}{1}

\begin{tabular}{ccc}
GDP & Gov. Spending & GDP Deflator \\

\includegraphics[
    width=0.32\textwidth,
    trim=1mm 1mm 1mm 1mm,
    clip
]{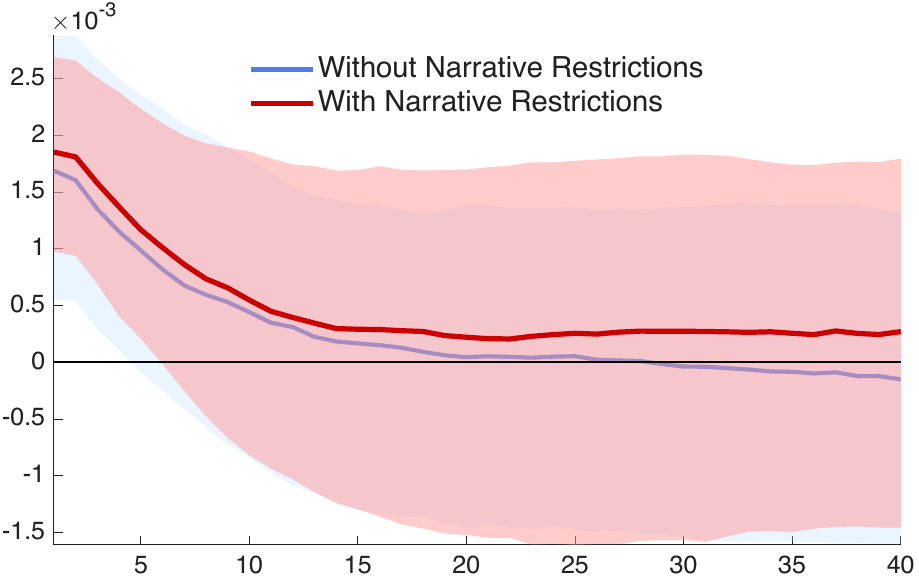}
&
\includegraphics[
    width=0.32\textwidth,
    trim=1mm 1mm 1mm 1mm,
    clip
]{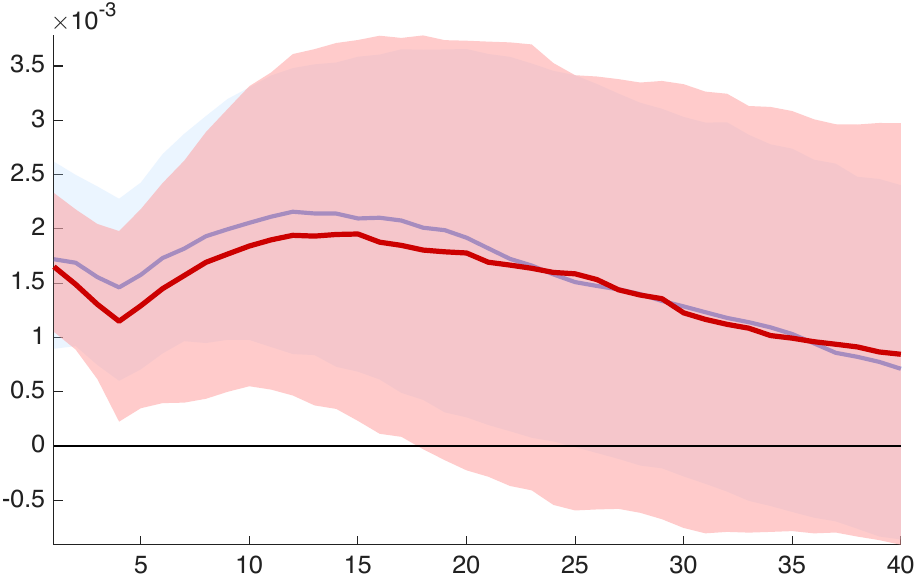}
&
\includegraphics[
    width=0.32\textwidth,
    trim=1mm 1mm 1mm 1mm,
    clip
]{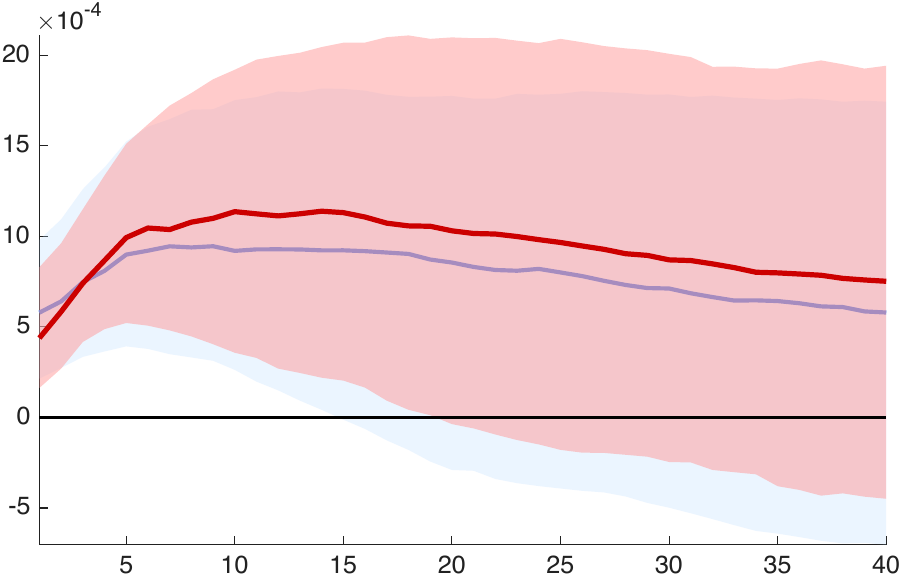}

\\[0.4cm]

CPI & Unemployment & Disposable Income \\

\includegraphics[
    width=0.32\textwidth,
    trim=1mm 1mm 1mm 1mm,
    clip
]{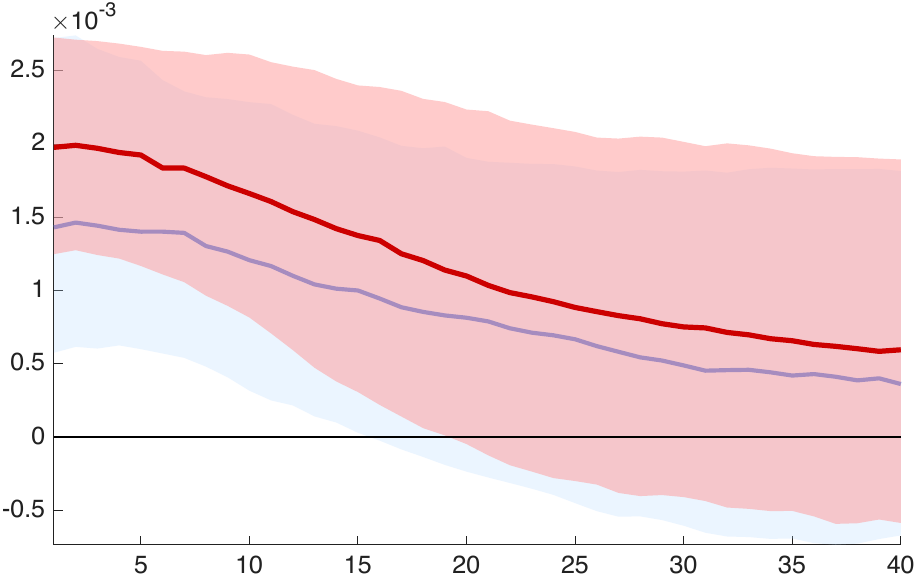}
&
\includegraphics[
    width=0.32\textwidth,
    trim=1mm 1mm 1mm 1mm,
    clip
]{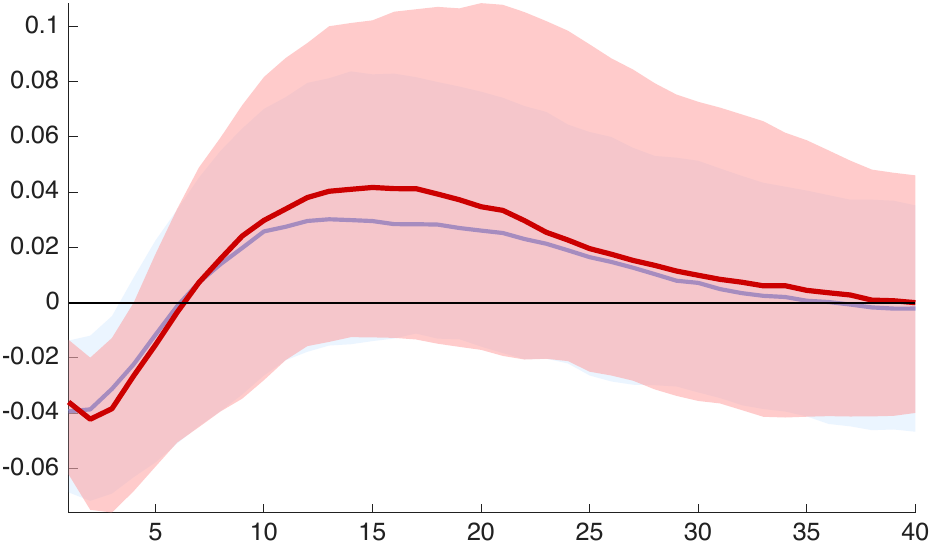}
&
\includegraphics[
    width=0.32\textwidth,
    trim=1mm 1mm 1mm 1mm,
    clip
]{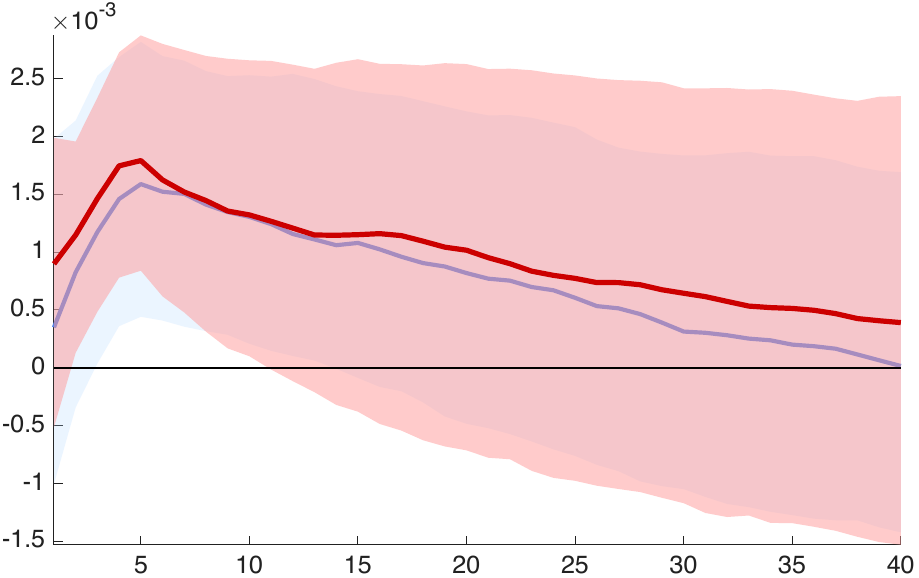}

\end{tabular}
\vspace{-0.25cm} 
\caption{Selected IRFs to Expansionary Government Spending Shock.} 
\label{fig2_government_shock}
\vspace{0.05cm} 
\begin{minipage}{0.98\textwidth}
\footnotesize
\textit{Notes:} Impulse responses of selected economic indicators.
Blue: 68\% credible sets (no narrative restrictions).
Red: median and 68\% bands (with narrative restrictions).
\end{minipage}
\end{figure}

\begin{figure}[H]
\centering
\setlength{\tabcolsep}{2pt}
\renewcommand{\arraystretch}{1}

\begin{tabular}{ccc}
GDP & Investment & CPI \\

\includegraphics[
    width=0.32\textwidth,
    trim=1mm 1mm 1mm 1mm,
    clip
]{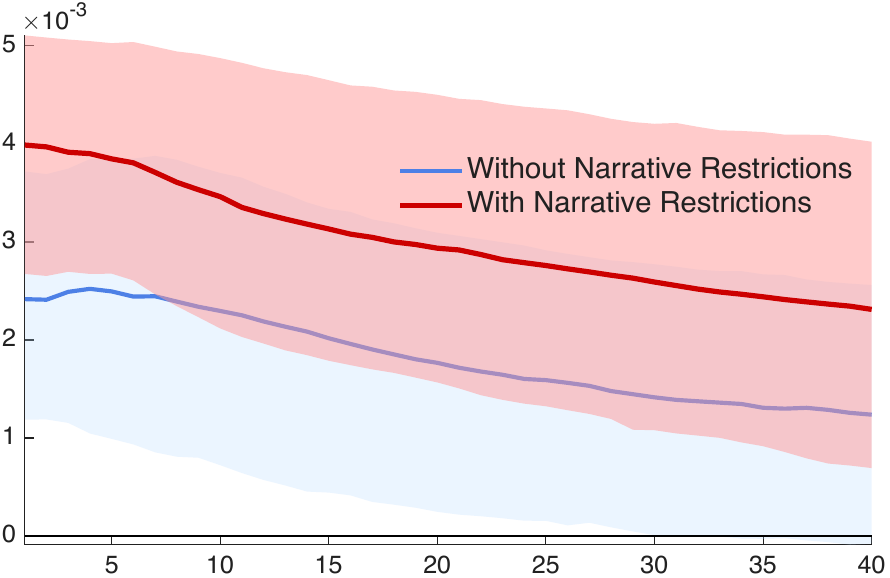}
&
\includegraphics[
    width=0.32\textwidth,
    trim=1mm 1mm 1mm 1mm,
    clip
]{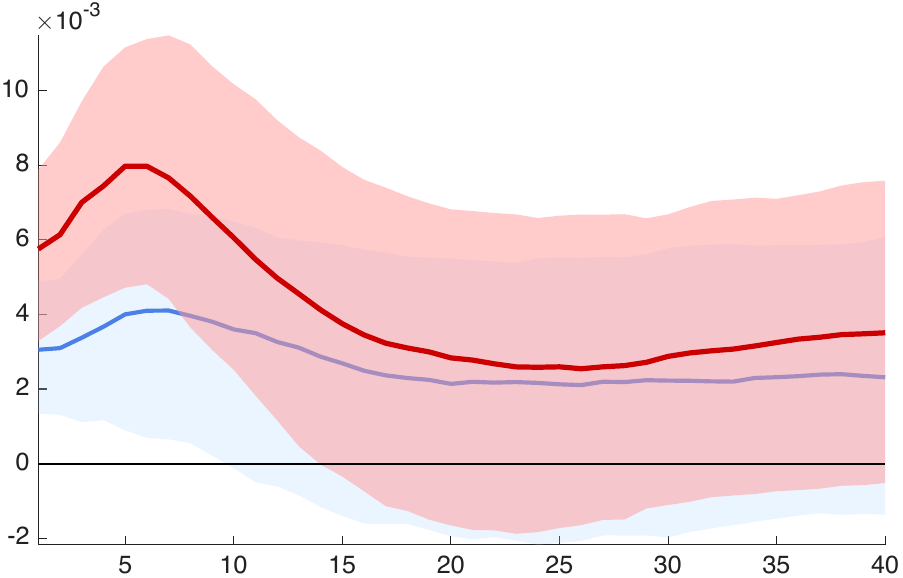}
&
\includegraphics[
    width=0.32\textwidth,
    trim=1mm 1mm 1mm 1mm,
    clip
]{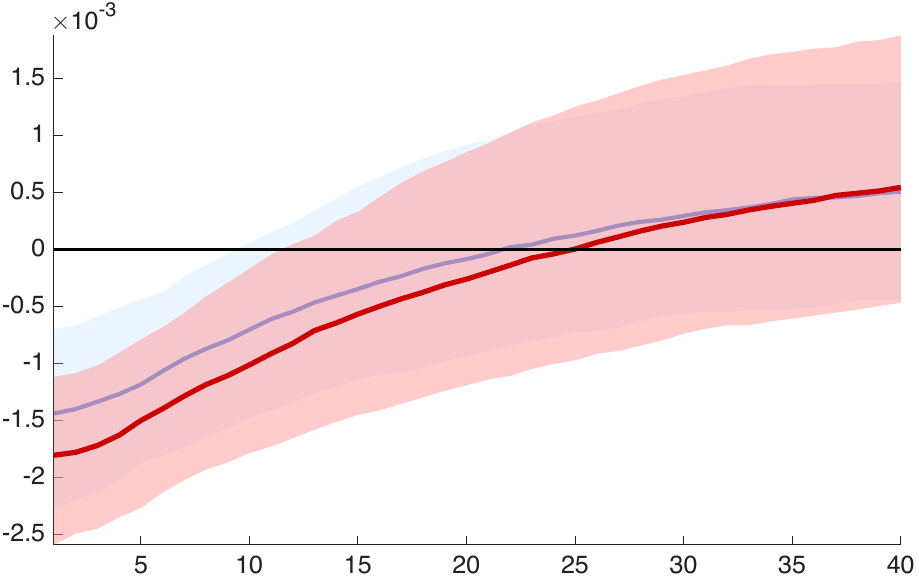}

\\[0.4cm]

Unemployment & Disposable Income & S\&P 500 \\

\includegraphics[
    width=0.32\textwidth,
    trim=1mm 1mm 1mm 1mm,
    clip
]{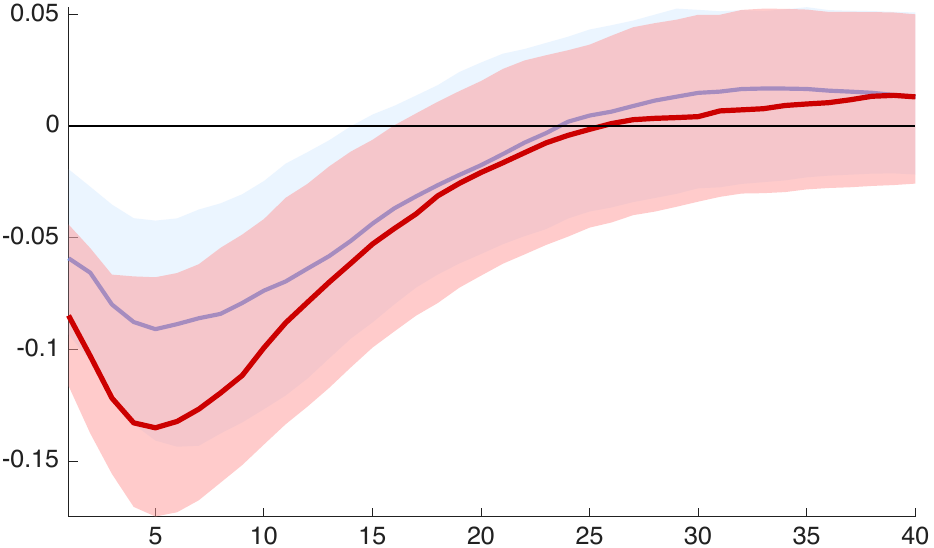}
&
\includegraphics[
    width=0.32\textwidth,
    trim=1mm 1mm 1mm 1mm,
    clip
]{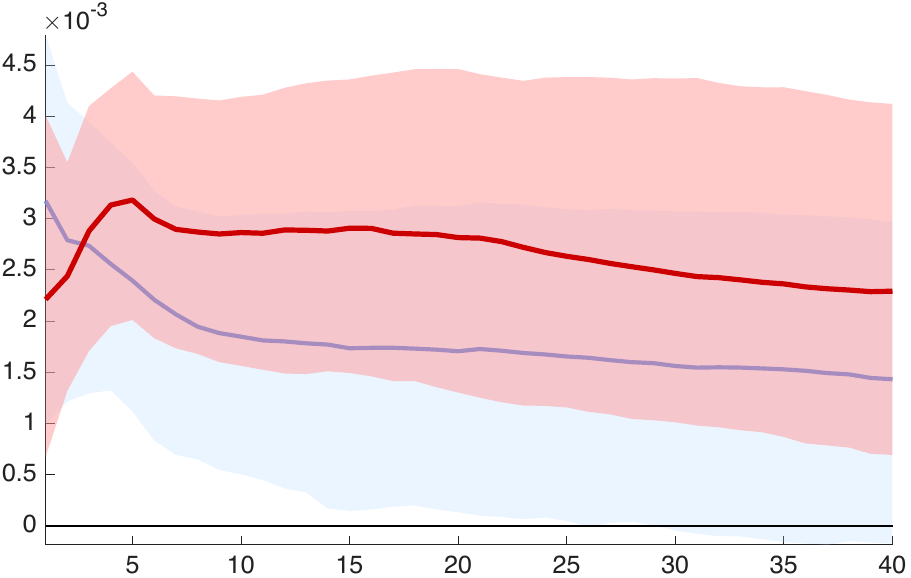}
&
\includegraphics[
    width=0.32\textwidth,
    trim=1mm 1mm 1mm 1mm,
    clip
]{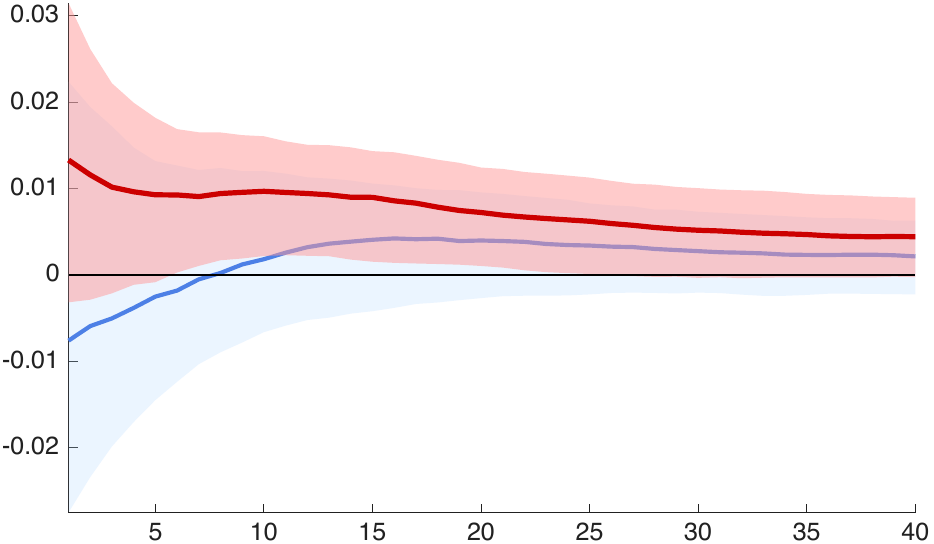}

\end{tabular}
\vspace{-0.25cm} 
\caption{Selected IRFs to Favorable Oil Supply Shock.} 
\label{fig2_oil_shock}
\vspace{0.05cm} 
\begin{minipage}{0.98\textwidth}
\footnotesize
\textit{Notes:} Impulse responses of selected economic indicators.
Blue: 68\% credible sets (no narrative restrictions).
Red: median and 68\% bands (with narrative restrictions).
\end{minipage}
\end{figure}

\begin{figure}[H]
\centering
\setlength{\tabcolsep}{2pt}
\renewcommand{\arraystretch}{1}

\begin{tabular}{ccc}
Demand Shock in 2006Q1 & Fin. Shock in 2007Q3 & Gov. Shock in 1980Q1 \\

\includegraphics[
    width=0.32\textwidth,
    trim=1mm 1mm 1mm 1mm,
    clip
]{"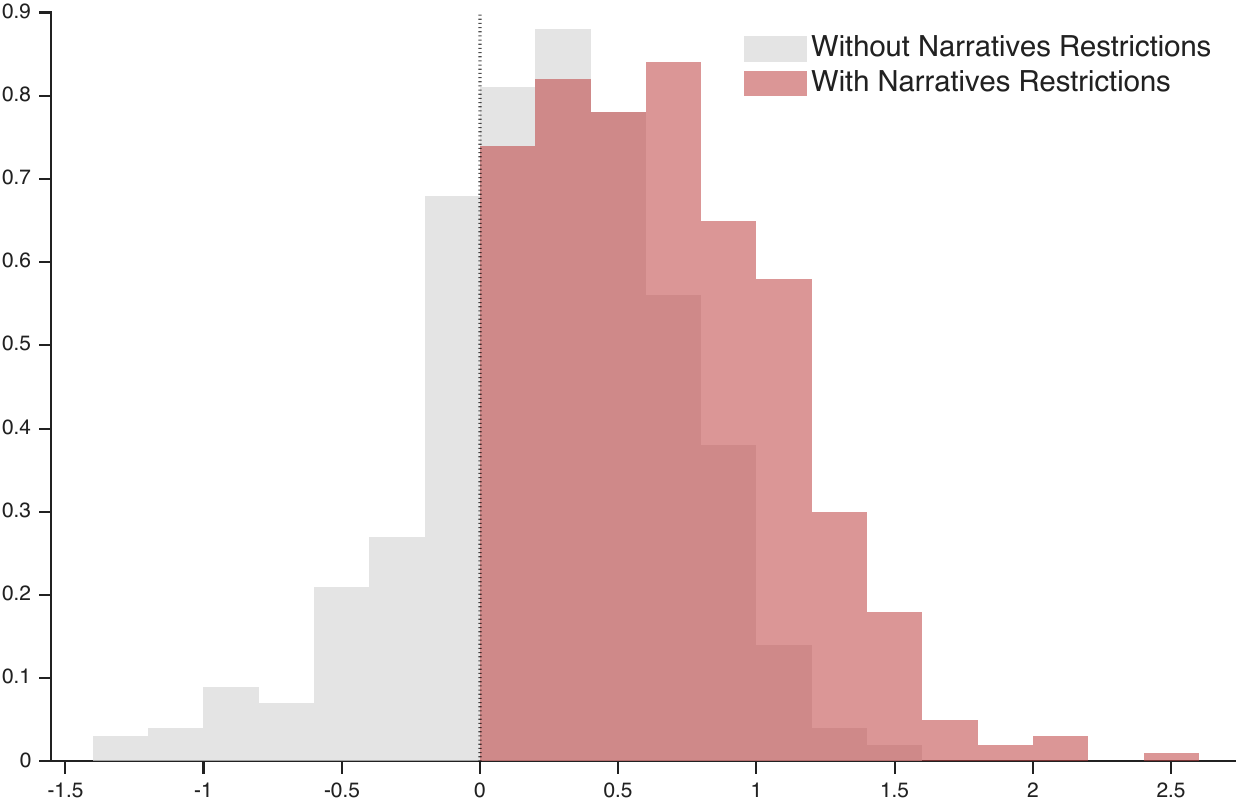"}
&
\includegraphics[
    width=0.32\textwidth,
    trim=1mm 1mm 1mm 1mm,
    clip
]{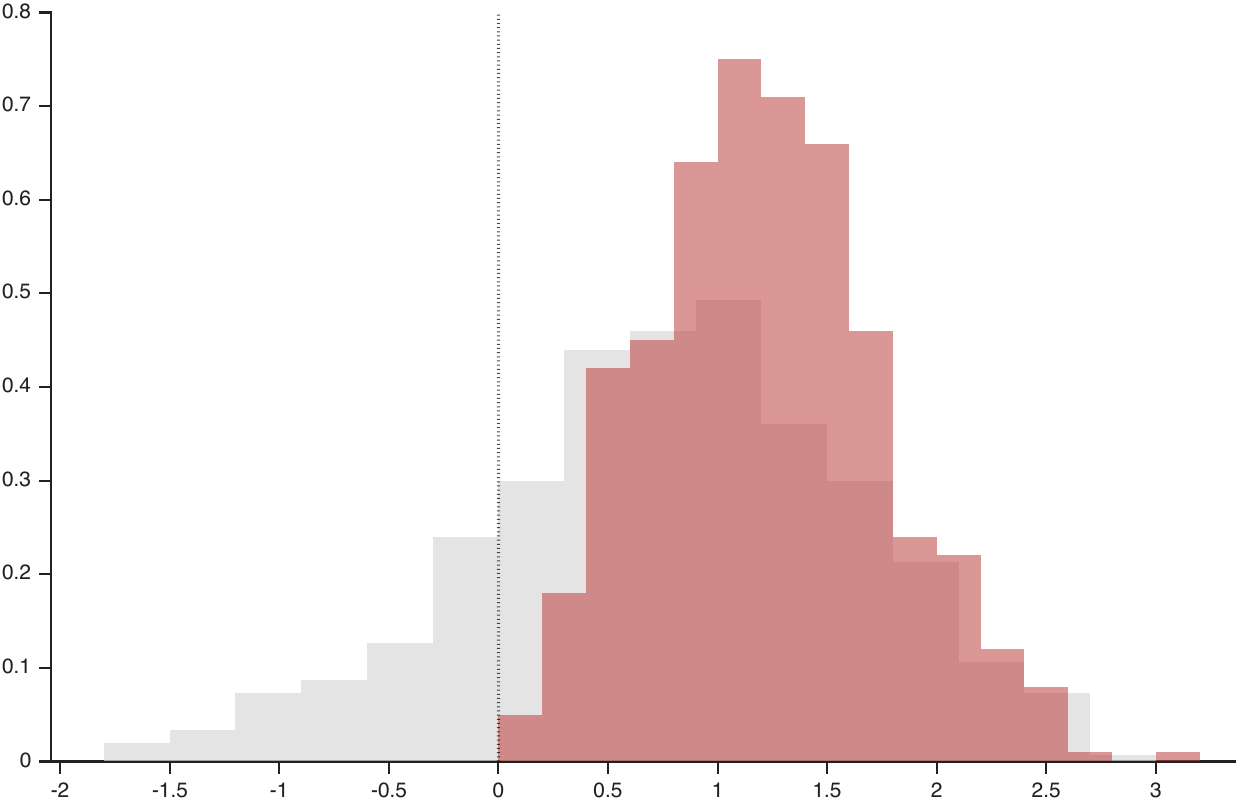}
&
\includegraphics[
    width=0.32\textwidth,
    trim=1mm 1mm 1mm 1mm,
    clip
]{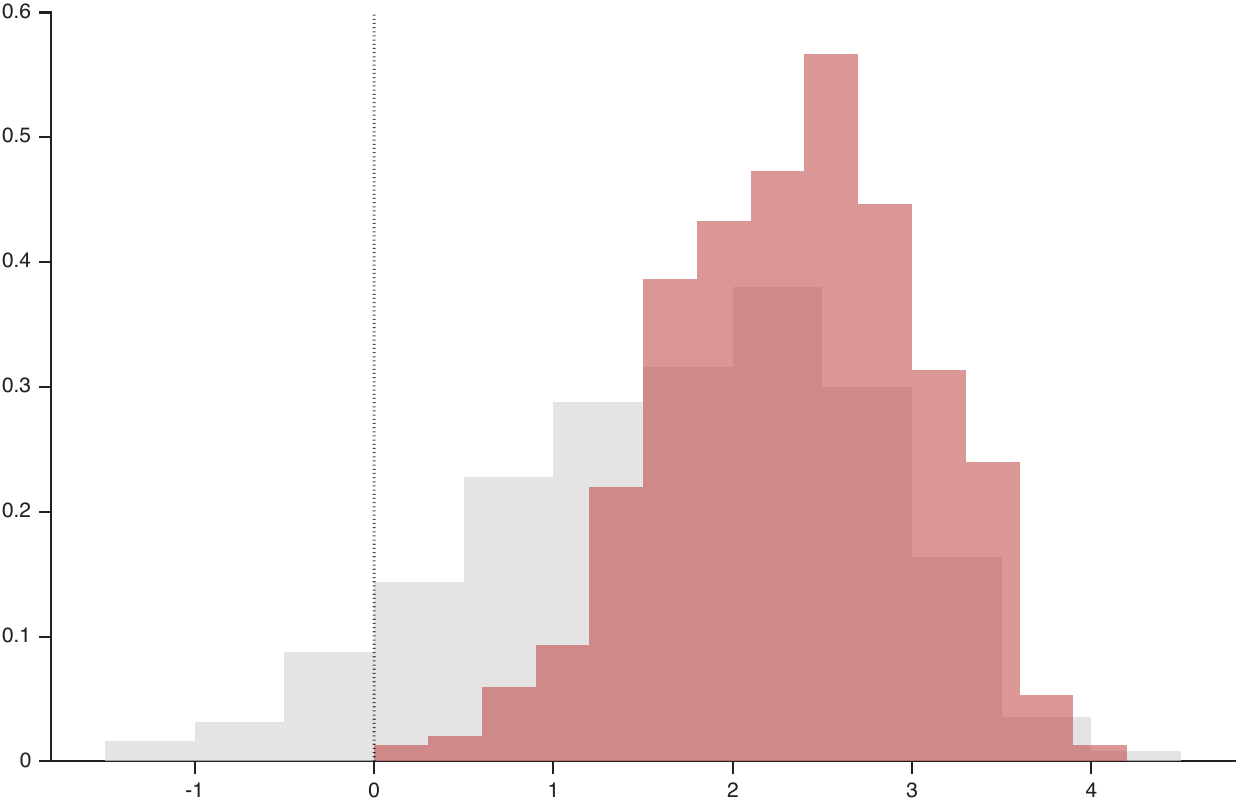}

\\[0.4cm]

Gov. Shock in 2001Q3 & Oil Shock in 1990Q3 &Oil Shock in 2002Q4 \\

\includegraphics[
    width=0.32\textwidth,
    trim=1mm 1mm 1mm 1mm,
    clip
]{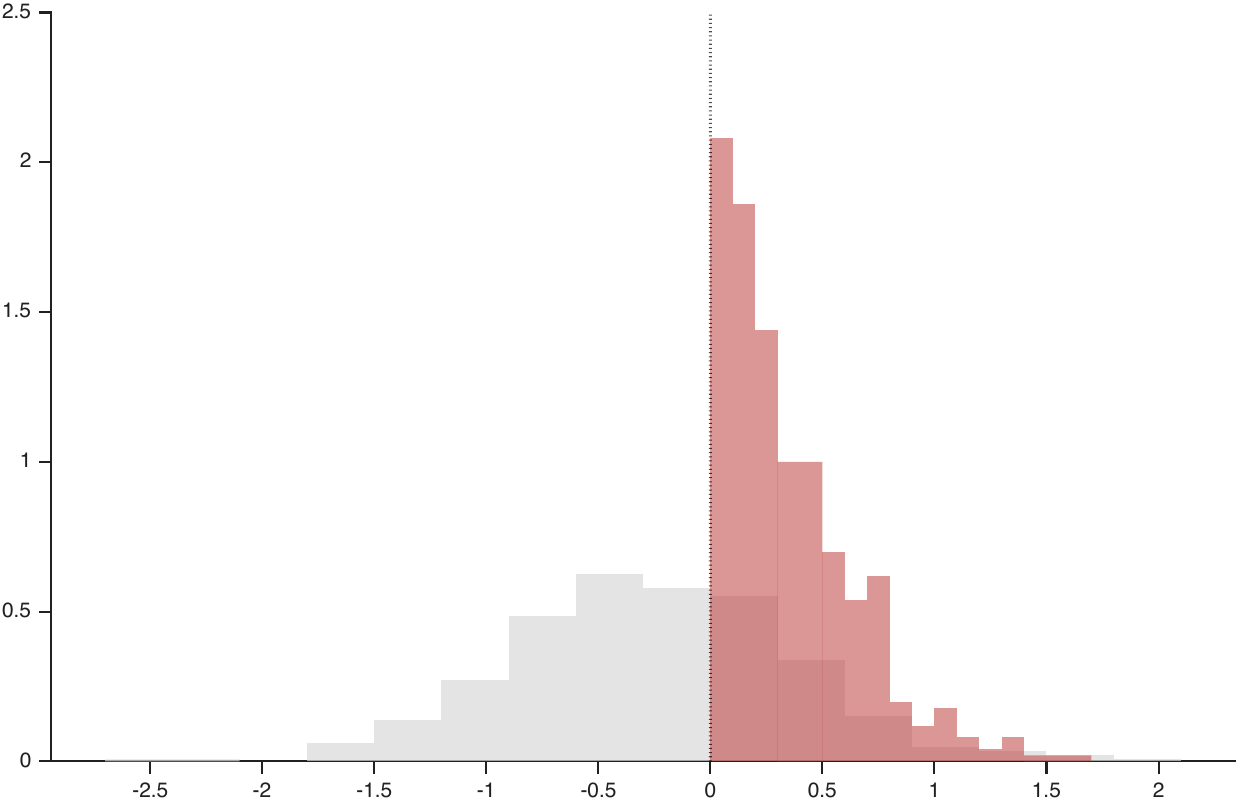}
&
\includegraphics[
    width=0.32\textwidth,
    trim=1mm 1mm 1mm 1mm,
    clip
]{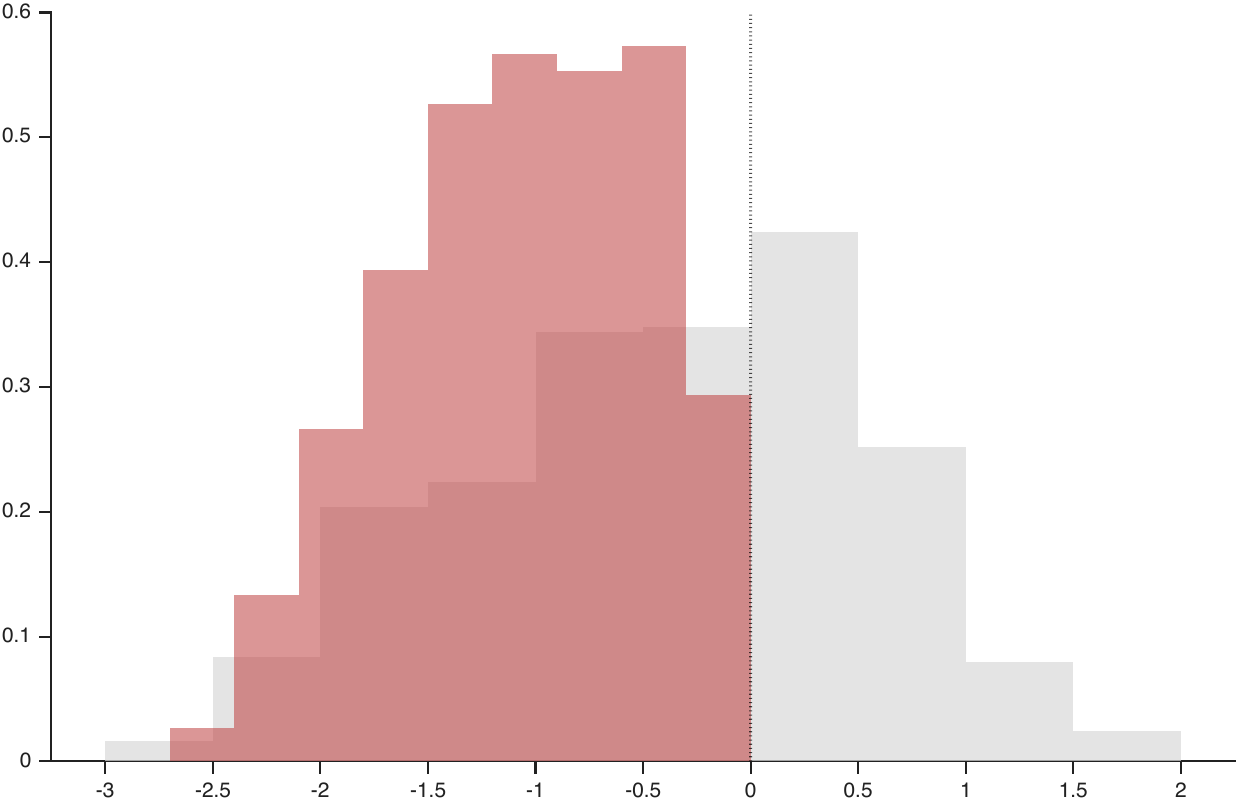}
&
\includegraphics[
    width=0.32\textwidth,
    trim=1mm 1mm 1mm 1mm,
    clip
]{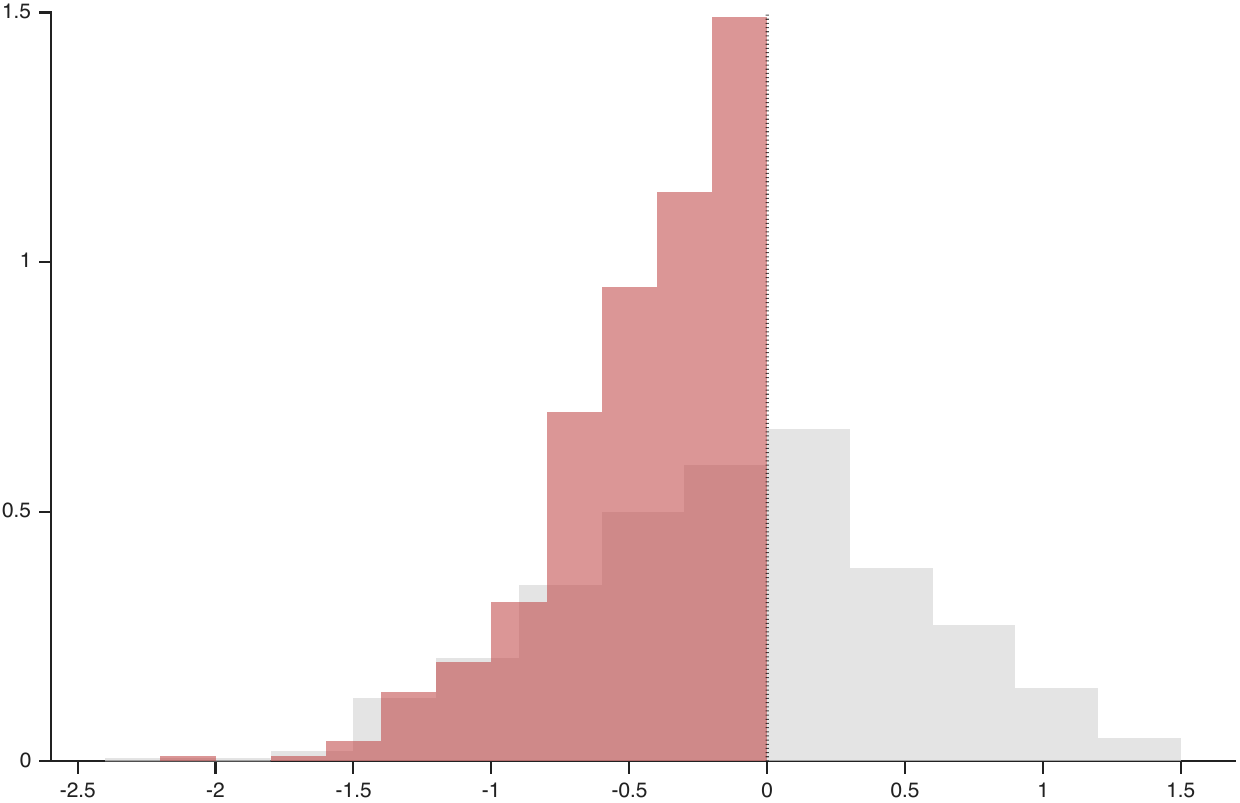}

\end{tabular}
\vspace{-0.25cm} 
\caption{Selected Posterior Distributions of Selected Shocks.} 
\label{fig_shock_distribution2}
\vspace{0.05cm} 
\begin{minipage}{0.98\textwidth}
\footnotesize
\textit{Notes:} The light gray histograms show the posterior distributions of the selected structural shocks in a given quarter for the extended model. The light red histograms show the corresponding posterior distributions for the model with narrative restrictions.
\end{minipage}
\end{figure}

\end{document}